\documentclass[prd,nofootinbib]{revtex4-1}

\usepackage{graphicx}
\usepackage{amsmath}
\usepackage{amssymb}
\usepackage{cleveref}
\usepackage{color}
\usepackage{float}
\usepackage{soul}
\usepackage{bbm}  
\usepackage{amsmath}                                                

\newcommand{\nc}{\newcommand}
\nc{\beq}{\begin{equation}}  \nc{\eeq}{\end{equation}}
\nc{\bea}{\begin{eqnarray}}  \nc{\eea}{\end{eqnarray}}
\nc{\baa}{\begin{array}}     \nc{\eaa}{\end{array}}
\nc{\bit}{\begin{itemize}}   \nc{\eit}{\end{itemize}}
\nc{\ben}{\begin{enumerate}} \nc{\een}{\end{enumerate}}
\nc{\bce}{\begin{center}}    \nc{\ece}{\end{center}}
\nc{\bpm}{\begin{pmatrix}}   \nc{\epm}{\end{pmatrix}}
\nc{\bvt}{\begin{verbatim}}  \nc{\evt}{\end{verbatim}}
\nc{\bal}{\begin{align}}
\def\mcr{\nonumber\\[6pt]}

\def\half{\frac12}	

\def\to{\rightarrow}

\def\gesim{\gtrsim}

\def\boldoverdot{\,{\raise6pt\hbox{\bf.}\!\!\!\!\>}}

\def\pp{{\bf p}}

\def\rr{{\bf r}}

\usepackage{bm}
\def\diag{\hbox{\diag}}

\def\ev{\hbox{eV}}

\def\gev{\hbox{GeV}}

\def\vevof#1{\left\langle #1 \right\rangle}

\def\doubleundertext#1{
{\undertext{\vphantom{y}#1}}\par\nobreak\vskip-\the\baselineskip\vskip4pt%
\undertext{\hbox to 2in{}}}
\def\inbox#1{\vbox{\hrule\hbox{\vrule\kern5pt
     \vbox{\kern5pt#1\kern5pt}\kern5pt\vrule}\hrule}}
\def\sqr#1#2{{\vcenter{\hrule height.#2pt
      \hbox{\vrule width.#2pt height#1pt \kern#1pt
         \vrule width.#2pt}
      \hrule height.#2pt}}}

\def\today{\ifcase\month\or
  January\or February\or March\or April\or May\or June\or
  July\or August\or September\or October\or November\or December\fi
  \space\number\day, \number\year}
\def\pmb#1{\setbox0=\hbox{#1}%
  \kern-.025em\copy0\kern-\wd0
  \kern.05em\copy0\kern-\wd0
  \kern-.025em\raise.0433em\box0 }

\def\pmbb#1{\setbox0=\hbox{#1}%
  \kern-.02em\copy0\kern-\wd0
  \kern.04em\copy0\kern-\wd0
  \kern-.02em\raise.03464em\box0 }
\def\inv#1{\frac1{#1}}
\def\sumprime_#1{\setbox0=\hbox{$\scriptstyle{#1}$}
  \setbox2=\hbox{$\displaystyle{\sum}$}
  \setbox4=\hbox{${}'\mathsurround=0pt$}
  \dimen0=.5\wd0 \advance\dimen0 by-.5\wd2
  \ifdim\dimen0>0pt
  \ifdim\dimen0>\wd4 \kern\wd4 \else\kern\dimen0\fi\fi
\mathop{{\sum}'}_{\kern-\wd4 #1}}
\def\vrot{v_{\tt rot}}
\def\vrotb{\overline{\vrot}}
\def\mpl{M_{\tt pl}}
\def\ub{\overline u}
\def\q{{\tt q}}
\def\w{X_B}
\def\rb{a}
\def\mbh{M_{\tt BH}}
\def\mb{M_{\tt B}}
\def\mdm{M_{\tt DM}}
\def\mhal{M_{\tt hal}}
\def\rhal{R_{\tt hal}}

\newcommand{\mnras}{MNRAS}

\begin{document}

\title{Ultra light Thomas-Fermi  Dark Matter}

\author{K. Pal}
\email{kuntal.pal@email.ucr.edu}
\author{L. V. Sales}
\email{lsales@ucr.edu}
\author{J. Wudka}
\email{jose.wudka@ucr.edu}

\affiliation{Department of Physics and Astronomy, UC Riverside, Riverside, CA 92521-0413, U.S.A}

\begin{abstract}
We investigate the viability of a simple dark matter (DM) model consisting of a single fermion in the context of galactic dynamics. We use a consistent approach that does not presume a particular DM density profile but instead requires that the DM+baryon system is in hydrostatic equilibrium. Using a phenomenological baryon density profile, the model then predicts the DM distribution with a core like behavior close to the galactic center. The presence of supermassive black holes (SMBHs) in the center of large galaxies arises naturally in this framework. Using data from a set of large elliptical and spiral galaxies, and from a small set of dwarf galaxies, we find that the model can explain most of the bulk galactic properties, as well as some of the features observed in the rotation curves, provided the DM mass is in the $\mathcal{O}(50\,\ev)$ range. More precise tests of the model require better modeling of the baryon profile and better control on the uncertainties in the data.
\end{abstract}

\maketitle

\section{Introduction}
\label{sec:introduction}

The nature of dark matter (DM) remains one of the most pressing  questions in modern cosmology and astrophysics; despite enormous theoretical and observational/experimental efforts, no definite DM candidate, or even paradigm for the dark sector, has been generally accepted. Direct probes of the dark sector, such as the direct detection experiments \cite{aprile2018dark, aguilar2017first, crisler2018sensei, angloher2017results} and collider searches \cite{boveia2018dark, Kahlhoefer:2017dnp}, have placed only limits on some of the interactions of dark particles. Cosmological and astrophysical observations have placed  complementary constraints, such as those derived form the relic abundance requirement \cite{Steigman:2012nb}, and the need to address the core-cusp problem \cite{Moore:1999gc} in the DM galactic distribution. For this last problem a popular approach has been to assume that the dark sector has appropriately strong, velocity-dependent self-interactions \cite{Spergel:1999mh}. An alternative idea~\footnote{The possibility that DM consists of ultra-light bosons that form a Bose-Einstein condensate on galactic scales has also been studied \cite{Boehmer:2007um, Alexander:2016glq} as a way of addressing the cusp problem.} is to assume that the DM is composed of fermions \cite{domcke2015dwarf, randall2017cores,destri2013fermionic}, and to ascribe the absence of a cusp to the exclusion principle; in this paper we investigate in some detail the viability of this last possibility.

Qualitatively speaking the possibility that the Pauli principle is responsible for the smooth DM profile at the galactic cores can be realized only for sufficiently light fermionic DM: only if the wavelength of such fermions is large enough can we expect the exclusion principle to be effective over distances typical of galaxies. This type of DM would be light; in fact, we will show below that the model provides reasonable results for masses $ \sim 50 \,\ev$, consistent with qualitative arguments~\cite{domcke2015dwarf}. Such light DM candidates could not have been in thermal equilibrium during the big-bang nucleosynthesis and large scale structure formation epochs \cite{Berezhiani:1995am, Feng:2008mu, Dodelson:1993je, Shi:1998km, Dodelson:2005tp, Boyarsky:2018tvu, Dasgupta:2013zpn}. This can be achieved by assuming the DM fermions carry a conserved charge, under which all standard model (SM) particles are neutral, in which case there are no renormalizable couplings between the DM fermions and the SM~\footnote{There are, of course, non-renormalizable couplings, but these are proportional to inverse powers of some scale -- the scale of the (heavy) physics that mediates such interactions. We assume that such scale is sufficiently large to ensure absence of SM-DM equilibrium.}. In this situation most constraints are easily met, with the exception of the relic abundance, for which existing approaches \cite{Dasgupta:2013zpn} can be adapted. Alternatively (though this is less attractive), the relic abundance can be ascribed to some primordial abundance generated in the very early universe by a yet-unknown mechanism. In this paper, however, we concentrate on galactic dynamics -- cosmological considerations lie outside the  scope of our investigation.

In the calculations below we obtain the DM distribution  assuming only {\it(i)} hydrostatic equilibrium, {\it(ii)} non-interacting and isothermal DM, {\it(iii)} asymptotically flat rotation cuves, and {\it(iv)} a given baryon density. More specifically, we do not make any assumptions about the shape of the DM distribution or its degree of degeneracy, which differs from the approach used in several related calculations that have appeared in  in the literature \cite{burkert1995structure, salucci2007, di2018phase,domcke2015dwarf}. One additional salient trait of this model is that it generally requires the presence of a super-massive black hole (SMBH) at the galactic center, though in special cases it can also accommodate galactic configurations without a SMBH.

An interesting argument found in the literature \cite{burkert1995structure, salucci2007, di2018phase,domcke2015dwarf}, based on the requirements that the assumed DM profile is consistent with the observational features (core size, velocity dispersion etc.) or merely from the DM phase space distribution \cite{Tremaine:1979we}~\footnote{Other lower bounds can be derived form the relic density constraint \cite{randall2017cores}, which we do not consider here.}, leads to a lower-bound constraint on the mass of the DM candidate. Our calculations do not generate this type of constraint because we make no {\it a-priori} assumptions about the DM distribution; in fact, we obtain consistent values as low as $ \sim 20  $ eV (cf. Sect. \ref{sec:large}).  In contrast, we do obtain an {\em upper} bound for the DM mass that depends on the asymptotic value of the rotation velocity and the mass of the SMBH (if no black hole is present the bound is trivial).

The rest of the paper is organized as follows. The equilibrium of the DM+baryon system is discussed in the next section; we then apply  the results to spherically-symmetric configurations (Sec. \ref{sec:SSS}). In Sec. \ref{sec:applications} we compare the model predictions with observational data for specific galaxies and obtain the DM mass values consistent with these observations. Conclusions are presented in Sec. \ref{sec:conclusions}, while some details of the data we used are provided in the Appendix.

\section{Equilibrium equations}
\label{sec:equil.eq}
As indicated above, we will investigate the viability of a Fermi-Dirac gas as a galactic DM candidate; we will assume that the gas is in local equilibrium, and that its self-interactions can be neglected. Additionally, we also assume the gas is non-relativistic, which we will justify {\it a posteriori}. In this case the hydrostatic stability of a small volume of the DM gas requires
\beq
m n \nabla \Phi + \nabla P =0 \,,
\label{eq:P}
\eeq
where $m $ is the DM mass, $n$ the density of the gas, $P$ its pressure, and $ \Phi $ the gravitational potential. Using  the standard thermodynamic relation $ n \, d\mu = dP - s \, dT $, where $ \mu $ is the chemical potential, $T$ the temperature and $s$ the entropy (volume) density of the gas, it follows that
\beq
\nabla(m \Phi + \mu) + \frac sn \nabla T=0 \,.
\eeq
We will assume that $T$ is constant throughout the gas, in which case 
\beq
m \Phi + \mu = E_0= \mbox{constant}.
\label{eq:energy}
\eeq 
The value of $E_0 $ will be discussed below.

Using \cref{eq:energy} in the Poisson equation for $ \Phi $  gives
\beq
\nabla^2 \mu = - \frac{4\pi m}{\mpl^2} \left( \rho_B +  m n \right)\,,
\label{eq:eom}
\eeq
where $ \mpl$ denotes the Planck mass~\footnote{We work in units where $k_B=  \hbar = c = 1 $, where $ k_B$ is Boltzmann's constant}, $ \rho_B $ is the baryon mass density, and  $n$  the DM number density (as noted previously); explicitly
\beq
n = - \frac 2{\lambda^3} \mbox{Li}_{3/2}\left( -e^{\mu/T} \right)\, ;\quad \lambda = \sqrt{\frac{2\pi}{m T}}\,,
\label{eq: n}
\eeq
where Li denotes the standard polylogarithm function and $\lambda $ is the thermal wavelength; the factor of 2 is due to the spin degrees of freedom.

Using standard expressions for the ideal Fermi gas  the average DM velocity dispersion is given by
\beq
\sigma^2_{\text{\tiny DM}} = \inv3\vevof{v^2}  = \frac P{m\, n} \,; \quad P = - \frac {2T}{\lambda^3} \mbox{Li}_{5/2}\left( -e^{\mu/T} \right)\,.
\label{eq:sigma}
\eeq

Within our model introduced above, the structure of the galaxy is determined by the solution to equation \cref{eq:eom} with appropriate boundary conditions. To do this our strategy will be to choose an analytic parameterization for $ \rho_B $ consistent with observations, and impose boundary conditions at large distances from the galactic center which lead to the flat rotation curves; from this $ \mu(\rr)$ can be obtained. The solution will depend on the parameters in $ \rho_B$, the DM mass $m$ and the asymptotic rotation velocity $\vrotb$. 

The idea of constraining the DM mass using the phase space density evolution was first suggested by Tremaine and Gunn (TG)~\cite{Tremaine:1979we}. In their seminal approach the DM halo is assumed to be an isothermal classical ideal gas in hydrostatic equilibrium, with a phase space distribution of the form $ f(\pp,\rr)= n(r) \exp[ - p^2/(2 m^2 \sigma^2) ]$, where $ n(r) = n_0/r^2 $. The exclusion principle then requires $ f(0,\rr) < 1 $, which leads to the lower bound  $m^4 \gesim 0.004  \mpl^2/(\sigma r^2)$. This bound then follows from a consistency requirement associated with the adopted form of $f$.  The Milky Way dwarf spheroidal satellites, due to their high DM density, allow a simple and robust application of the TG bound (see for eg. \cite{boyarsky2009lower,randall2017cores})  obtaining, for example, $ m \gesim 70$ eV using Fornax \cite{randall2017cores}, though uncertainty in the DM core radius limits somewhat the reliability of this bound~\footnote{A large core size cannot be ruled out \cite{di2018phase}, while relaxing the dependence of the DM halo core radius on the stellar component and  marginalizing the unknown stellar velocity dispersion anisotropy lead to mass bounds as low as $20$ eV  \cite{di2018phase} (though such large haloes are unrealistic and would be at odds with their lifetime due to  dynamical friction effects within the Milky Way).}. 

In contrast to these assumptions, we use the Fermi-Dirac distribution $ f(\pp,\rr) = \{ \exp [ p^2/(2m T) - \mu(r)/T ] + 1 \}^{-1}$ that, $i)$ automatically satisfies the exclusion principle constraint,  $ii)$ does not factorize into a product of space and momentum functions and $iii)$ leads to a singular $n$ only when a central SMBH is present. In our approach the DM density profile is determined by the baryon distribution by solving  \cref{eq:eom}; we make no assumptions about the the DM core radius or the DM distribution in general. In particular, the degree of degeneracy of the fermion distribution function follows from the behavior of $ \mu(\rr) $;  we will see below that the DM approximates a classical Maxwell-Boltzmann gas far from the bulge and that its quantum nature only becomes important near the galactic center, leading to a core-like profile. Despite these differences, we observe that the values we obtain for $m$ (see below) are roughly consistent with the bounds based on the extended TG approach, especially for smaller dwarf galaxies with higher DM density.

The value of $ \mu $ at the origin will be of interest in interpreting the solutions to \cref{eq:eom}. If $ \mu(\rr\to0) \to + \infty $ then $ \phi \to - \infty $, which, as we will show, corresponds to a point-like mass at origin, a black hole \footnote{This scenario was recently considered in \cite{de2017warmbh} with completely different boundary conditions, without baryons and DM mass in the keV range.}. In these cases, the DM density exhibits a cusp at the origin, but for realistic parameters this cusp appears only in the immediate vicinity of the black hole. Outside this region the DM density has a core-like profile. Solutions for which $ \mu(0) $ is finite corresponds to galaxies where no central black hole is present and exhibit `pure' core-like DM densities. The remaining possibility,  $ \phi(\rr\to0)\to + \infty $ describe the unphysical situation of a repulsive point-like object.

\section{Spherically symmetric solutions}
\label{sec:SSS}

In the following section we will adopt the simplifying assumption that all quantities depend only on $r = |\rr|$; this is a reasonable assumption for ellipticals, but is problematic for spiral galaxies. We will comment on this when we apply our formalism to specific cases.

It  proves convenient to define   $ \ub$ and $x$ by
\beq
x = \frac r A\,, \quad \frac{\ub(x)} x = \frac\mu T\,; \quad A = \sqrt{\frac{T \mpl^2 \lambda^3}{8\pi m^2}}\,,
\label{eq:ubdef}
\eeq
while the baryon density can be written in the form
\beq
\rho_B = \frac{\mb}{\left(\frac43\pi  \rb ^3 \right)} F(r/ \rb )\,,
\label{eq:B-density}
\eeq
where $\mb$ is the total bulge mass and $  \rb $ denotes the scale radius which can be obtained from the effective radius using the explicit form of the baryonic profile function $F$; $ \rho_B $ will be negligible for $ r \gg  \rb $. The normalization for $F$ is taken to be
\beq
 \int_0^\infty dy\,y^2 F(y) = \inv3\,.
\eeq

With these definitions \cref{eq:eom} becomes (a prime denotes an $x$ derivative)
\beq
\ub'' =  x \mbox{Li}_{3/2}\left( - e^{\ub/x} \right) -\q\, x F(x/\w)\,, \qquad \w =  \rb /A\,,\quad \q=\frac{3\mb \lambda^3}{8\pi m  \rb ^3 } \,.
\label{eq:ueom}
\eeq
For most of the examples we consider $ \w \lesssim 1 $.

Far from the galactic center $ \rho_B $ can be neglected and  the gas density will be small enough so that $ P = n T $ and  Li$_{3/2}(-z) \simeq -z $. In this region a `test' object in a circular orbit of radius $r$ will have velocity  
$ \vrot(r)$ determined by
\beq
 \vrot^2(r) = \frac{ M_{\tt tot}(r)}{\mpl^2 r }\,,
 \label{eq:vrot}
\eeq
where $ M_{\tt tot}$ is the total mass ($\mbh$ + $\mb$ + $\mdm$) inside radius $r$. At large distances $\vrot(r)$ will approach an $r$-independent value $ \vrotb$ provided $ M_{\tt tot}(r) \propto r$, which requires $ n \sim 1/r^2$ (since the dark component dominates in the asymptotic region).  This then implies $ \ub = x \ln(b/x^2) $ for some constant $b$; substituting in $ \ub'' \simeq - x \exp(\ub/x) $ gives $ b=2$:
\beq
\ub \to  x \ln \left( \frac2{x^2}\right)\,, \quad x \gg \w.
\label{eq:uas}
\eeq
The numerical solutions approach the asymptotic expression in \cref{eq:uas}  for $ x \gesim 1$.

Using the asymptotic expressions it follows that $ M_{\tt tot}(r) \simeq (16\pi A^2/\lambda^3)m r $, whence \cref{eq:vrot} gives
\beq
 T = \half m \vrotb^2 \,; \qquad \mbox{where}~~\vrot(r)\,  \stackrel{r \gg \rb}\longrightarrow \,\vrotb\,.
 \label{eq:T-vrot}
\eeq 
Comparing this with the expression \cref{eq:sigma} we find
\beq
\sigma^{}_{\text{\tiny DM}} = \frac\vrotb{\sqrt{2}} \,, \quad (r \gg  \rb ) \,;
\eeq
it also follows that $ \lambda = \sqrt{4 \pi}/( m \vrotb)$

We solve \cref{eq:ueom}  using \cref{eq:uas} and its $x$ derivative as boundary conditions. The solution~\footnote{For later convenience we explicitly display the dependence on the parameters $ \w$ and $\q$.} $ \ub(x;\w,\q)$ will then ensure that rotation curves are flat and is consistent with the chosen baryon profile. Note that in general $ \ub $ will not vanish at the origin, which implies the behavior
\beq
\Phi  \, \stackrel {r \to 0 }\longrightarrow \,  - \frac{A T}m \frac{u_0 }r\,, \quad u_0 = \ub(0;\w,\q) \,.
\label{eq:u_o}
\eeq
For $u_0>0 $ this corresponds to the field generated by a point mass 
\beq
 \mbh = \frac{AT \mpl^2}m\, u_0 = \left( \frac{\sqrt{\pi} \, \vrotb^3}8 \right)^{1/2} \frac{\mpl^3}{m^2} u_0
 \label{eq:mbh-u0}
 \eeq 
that we interpret as a black hole at the galactic center: in these cases the boundary conditions are consistent only if a black hole with this particular mass is present. For $u_0<0 $  the solution in  \cref{eq:u_o}  is unphysical, at least as far as classical non-relativistic configurations are concerned. These two regimes are separated by the  curve $ u_0 =0 $ in the $ \w - \q$ plane; solutions of this type correspond to galaxies without a central black hole.

For the examples that follow, we consider that the expression $ \ub(0;\w,\q)= u_0 $ is equivalent (to a good approximation) to the simple relation~\footnote{This was obtained numerically, not  derived rigorously using the properties of the solutions to \cref{eq:eom}.}
\beq
\ln \w = \nu(u_0) \ln\q + c(u_0)\,,
\label{eq:scal.rel}
\eeq
where the functions  $ \nu $ and $c$  depend on the form of $F$ in \cref{eq:B-density},  but are generally $\mathcal{O}(1) $. For the choices of $F$, and for $ u_0 $ not too close to zero, below they can be approximated by algebraic functions:
\beq
c(u_0) \sim \bar c_1 \sqrt{\bar c_2 - u_0}\,, \qquad \nu(u_0 ) \sim -\bar\nu_1 - \bar\nu_2 u_0^2 + \bar\nu_3 u_0^3\,,
\label{eq:c.nu}
\eeq
where $ \bar c_{1,2},\,\bar\nu_{1,2,3} $ are positive and $ O(1) $; values for several choices of $F$ are provided in the next section, see table \ref{fit.params}. The errors in using these expressions are below $ 10\% $, so they are useful for $ u_0 \gg 0.1$. Unfortunately, many cases of interest correspond to $ u_0 \lesssim 0.1 $, so in most results below we will not use \cref{eq:scal.rel,eq:c.nu}, opting instead for a high-precision numerical calculation.

 It is worth pointing out that  once the boundary conditions at large $r$ are imposed, $u_0 $ is determined by $ \w$ and $\q$, it is not a free parameter. Equivalently,  $ \mbh$ is determined by $m$ and $ \rho_B$, in particular, the presence (or absence) of a black-hole and its mass are not an additional assumption, but instead follow naturally from the choice of DM mass and baryon density profile. 

The relation \cref{eq:scal.rel} can be used to estimate the DM mass $m$ in terms of the galactic quantities $\mb,\, \rb$ and $\mbh$. Since $ c(u_0) $ in \cref{eq:c.nu} should be real, a necessary condition for $m$ to be real as well is $ u_0 < \bar c_2 $. This leads to the requirement:
\beq
m^2 <  \frac{\bar c_2}{(64/\pi)^{1/4}} \frac{\left( \mpl^2 \vrotb \right)^{3/2}}\mbh\stackrel{\bar c_2 =1.3}\sim \left( 180\, \ev \right)^2 \frac{ \left( 10^3 \vrotb \right)^{3/2}}{\mbh/\left(10^9 M_\odot\right)}\,;
\label{eq:con.con}
\eeq 
for most of the the specific examples studied below we find $ m \lesssim 100\,\ev$ (see Sec. \ref{sec:applications}).

To get an estimate of the values of the quantities involved, for $ m\sim 10\, \ev $ and $ \vrotb \sim 300 $km/s, $A\sim 20 $kpc and $ \mbh \sim 10^{11} u_0^2 M_\odot$, so that realistic situations will correspond to small values of $u_0 $ that will satisfy \cref{eq:con.con}. 

Since $ \vrotb \ll1 $ for all cases of interest, the gas temperature will be much smaller than its mass. In addition, $ \mu/m = \vrotb^2 \ub/(2x) $ (cf. \cref{eq:ubdef}), where we expect $ \ub \sim \mathcal{O}(1) $ (see Sec. \ref{sec:applications}) and $ \mu \ll m$; except perhaps in the immediate vicinity of the galactic center and even then only when $ \ub(0)\not=0$ (corresponding to $\mbh\not=0$). From this it follows that in general the Fermi gas will be  non-relativistic, as we assumed above.

We will define the halo (or virial) radius $ \rhal $ by the condition $ m n(\rhal) = 200 \times \rho_c $, where $ \rho_c  \simeq  4.21 \times 10^{-47} \gev^4 $ is the critical density of the Universe. For all cases considered here the density will take its asymptotic expression (corresponding to \cref{eq:uas}) at $ r = \rhal $, then we find
\beq
\rhal = \left(10^3 \vrotb \right)\times 240 \, \mbox{kpc} \,,
\label{eq:rhal}
\eeq
and depends only on $ \vrotb $; the galactic radius is then $\mathcal{O}(100\,\mbox{kpc})$.

Taking the zero of energy at infinity imposes the boundary condition $ \Phi(\rhal) = - \mhal/(  \mpl^2 \rhal) $ so that, using \cref{eq:energy} and \cref{eq:uas},
\beq
E_0 = T \ln \left( \frac{2A^2}{\rhal^2} \right) - \frac\mhal{\rhal \mpl^2}\,; \quad \mhal = \mb + 4\pi m \int_0^{\rhal} dr\, r^2 n(r) \,,
\eeq
so that $E_0 $ is then determined by the other parameters in the model.

\subsection{Sample calculation}
To illustrate the model presented above we consider a set of 3 hypothetical galaxies (cf. table \ref{examples}) for which we display some of the results derived from the calculations described above,  where the black-hole mass $ \mbh $ is calculated using \cref{eq:mbh-u0}. In this section we will assume $ m = 50 $ eV and use the Plummer profile $ F(y) = (1 + y^{2})^{-5/2}$ (again, for illustration purposes); note that the solution is independent of $a$ when $ \mb=0$.

\begin{table}
\caption{Sample galaxies}
$$
\begin{array}{|c|c|c||c|}
\hline
\text{Galaxy} & \mb/M_\odot & a\, (\text{kpc})& \mbh/M_\odot \cr
\hline
A & 0                   & --   & 8.5 \times 10^9 \cr \hline
B & 2.55 \times 10^{10} & 2.5  & 5.4 \times 10^7 \cr \hline
C & 2.55 \times 10^{10} & 3.25 & 2.8 \times 10^9 \cr  \hline
\end{array}
$$
\label{examples}
\end{table}

\begin{figure}[ht]
$$
\includegraphics[width=2.9in]{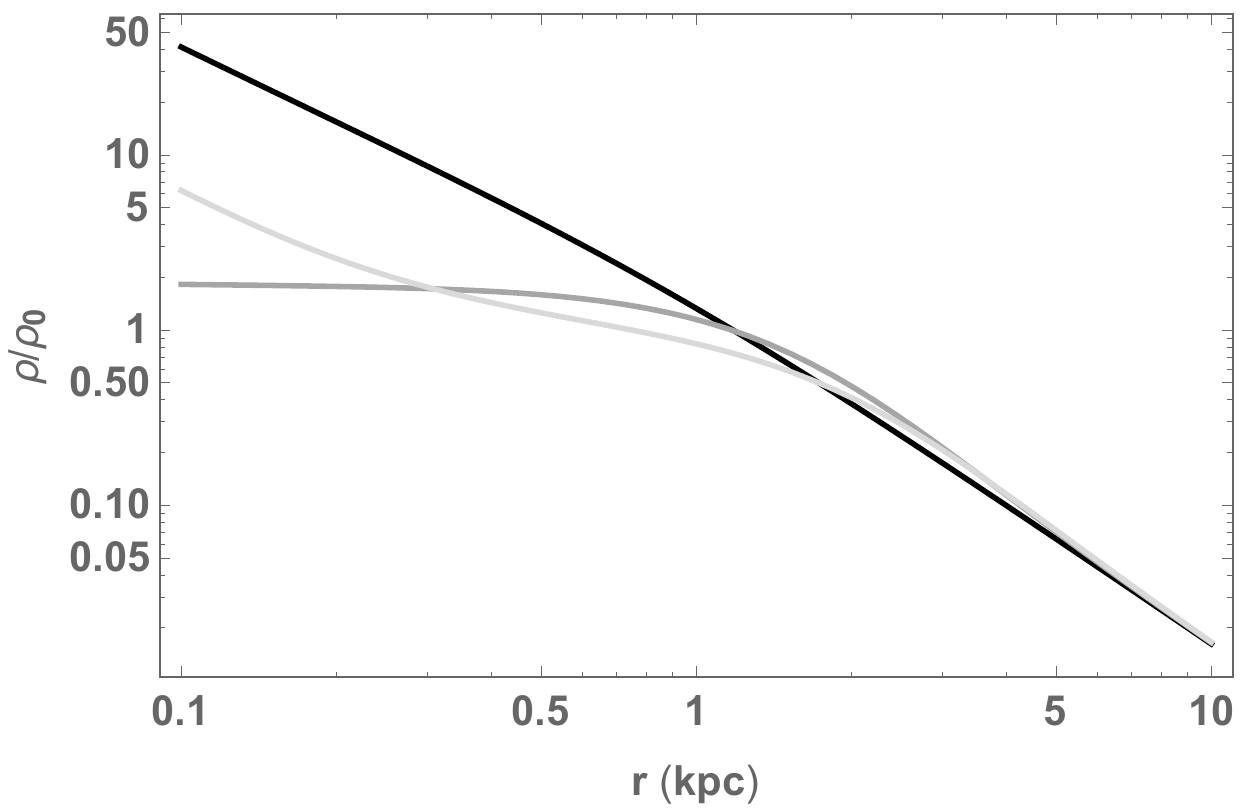} \qquad
\includegraphics[width=2.9in]{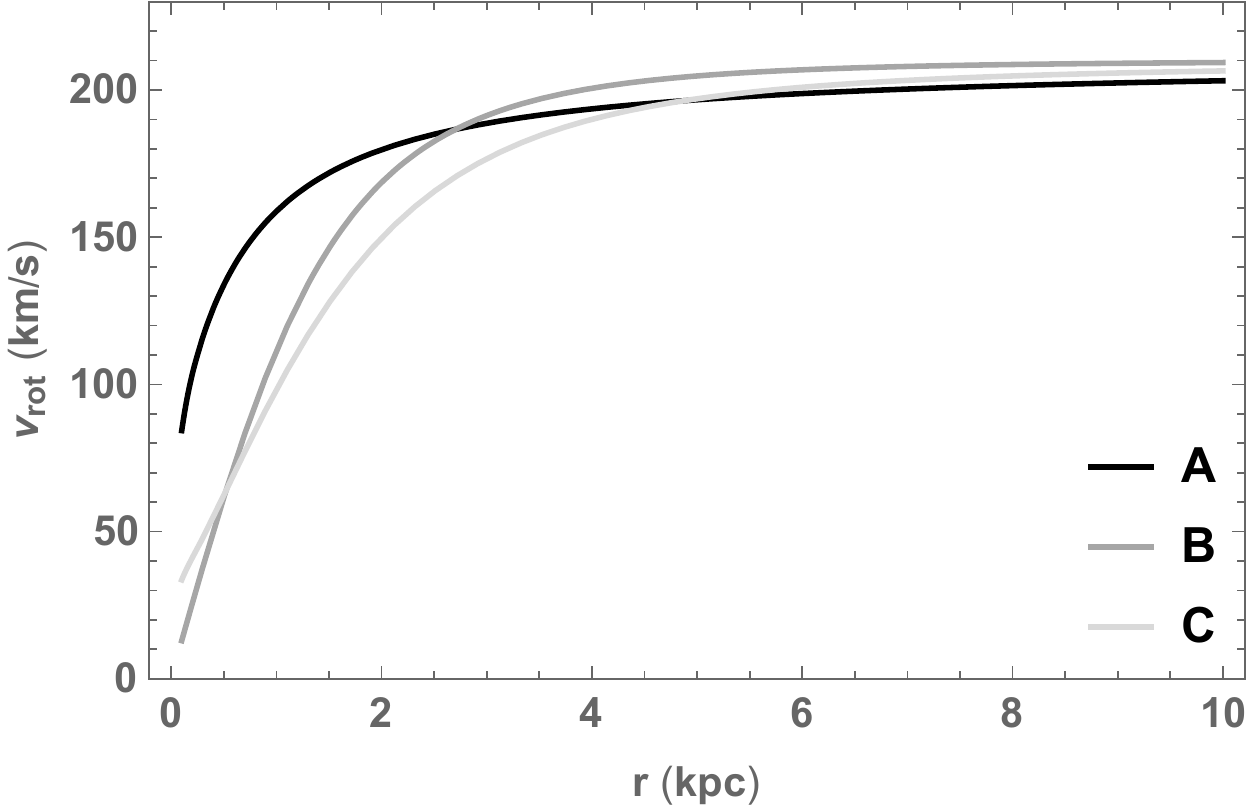} 
$$
	\caption{\footnotesize Density (left) and circular velocity (right) for the sample galaxies in table \ref{examples}, the black, dark-gray and light-gray curves correspond, respectively, to galaxies $A$, $B$ and $C$ ($\rho_0 = 2 m /\lambda^3$).}
\label{fig:example}
\end{figure}

All these galaxies have a halo radius (cf. \cref{eq:rhal}) $ \sim 300 $kpc. The total mass density and circular velocity \cref{eq:vrot} are plotted in Fig. \ref{fig:example}. Galaxy $A$ shows a density profile with no evidence for a core while a clear constant-density core develops in galaxies $B$ and $C$. Note that for the latter, the density increases again for $ r \lesssim 200 $ pc due to the relatively large central SMBH. Similarly, galaxy $B$ has a density increase only at very small radii, $ r \ll 100 $ pc, because of a smaller black hole at the galactic center.  The circular velocity profile is generally steepest for $A$, decreasing for $B$ and even more for $C$.

As shown by this exercise, the solution is very sensitive to the particular combination of size and galaxy mass ($\rb$ and $\mb$). For example, a change by a factor of 50 is predicted in  $ \mbh $ due to a relatively small ($\sim 40$\%) change in $\rb$, leading ultimately to quite different density profiles. While this can be considered a feature of the model, which is anticipated to have large predictive power, given the uncertainties that plague current astronomical measurements one may refrain from over-interpreting the results at such level of detail.

It is interesting to note that the case $ \mb=0 $ is universal, in the sense that the solution to \cref{eq:eom} with boundary conditions \cref{eq:uas} is unique and, in  particular, has $ u_0 \simeq 1.49 $: within this model configurations without a smooth baryon density are consistent with flat rotation curves only if they contain a central SMBH with mass $\sim (6 \times 10^6/m_{\tt eV})^2 M_\odot $ (see \cref{eq:mbh-u0}), where $ m_{\tt eV} $ is the DM mass in eV units.

\section{The TFDM model in specific galaxies.}
\label{sec:applications}

Given a spherically-symmetric galaxy with a known baryon density profile and a given black hole mass, the results of Sec. \ref{sec:SSS} predict a DM mass $m$. It is then important  to determine whether the {\em same} value of $m$ is obtained for different galaxies, as required for consistency. In this section we discuss this issue for a set of large galaxies (Sec. \ref{sec:large}) and then for a set of dwarf galaxies (Sec. \ref{sec:dwarf}). We note that we cannot expect a perfect agreement (that is, precisely the same $m$ in all cases), as we have ignored many of the details of the structure of the galaxies being considered (assuming, for example, spherical symmetry). We will be satisfied instead to see if the  values of $m$ derived for each galaxy cluster around a specific range.

\subsection{Large galaxies with SMBH}
\label{sec:large}

We will adopt the following three commonly used stellar density profiles (cf. \cref{eq:B-density}) \cite{plummer1911problem,hernquist1990analytical,jaffe1983simple} into our model.
\begin{align}
\label{eqn:profiles}
F(y) &=\frac{1}{(1 + y^{2})^{\frac{5}{2}}}  \qquad  \qquad \text{\text{(Plummer)}} \,, \mcr
F(y) &=\frac{2}{3y(1 + y)^{3}}  \qquad  \qquad  \textrm{\text{(Hernquist)}} \,, \mcr
F(y) &=\frac{1}{3y^2(1 + y)^{2}} \qquad  \qquad  \textrm{\text{(Jaffe)}} \,,
\end{align}
for which the parameters in the scaling relations \cref{eq:scal.rel,eq:c.nu} are provided in table \ref{fit.params}. We use different profiles in order to gauge the effect of baryon distribution on the DM mass in the set of galaxies that we study.

\begin{table}[!htb]
\caption{Fit Parameters}
\label{fit.params}
\vspace{-15pt}
$$
\begin{array}{|c|c|c|c|c|c|} 
 \hline
 \text{Baryon Profile} & \bar c_1 & \bar c _2 & \bar\nu_1 & \bar\nu_2 &\bar\nu_3 \cr
 \hline
  \text{Plummer} & 0.746 & 1.354 & 0.412 & 0.03 & 0.036\cr
 \text{Hernquist} & 0.931 & 1.404 & 0.434 & 0.08 & 0.069\cr
 \text{Jaffe} & 0.714 & 1.298  & 0.385 &  0.001 & 0.008\cr
 \hline
\end{array}
$$
\end{table}

We collected a dataset from several sources \cite{hu2009black, kormendy2013coevolution, lelli2017one, graham2016normal,sofue1999central,sofue2009unified} for a total of 60 galaxies, spanning a large range of Hubble types, and each of them containing a SMBH at their galactic center; details on data selection are  provided in appendix \ref{appendix:data}. Using the central values of $\mb,\, \mbh,\,\vrotb$, and $\rb$ provided in the above references, we calculate the DM mass $m$ for all the galaxies in this set~\footnote{To minimize inaccuracies, we do not use \cref{eq:scal.rel}, but  find $m$ by solving $ \ub(0;\w,\q)= u_0$ numerically.}. The results are shown in Table \ref{table:elliptical_galaxies} and Table \ref{table:spiral_galaxies} for elliptical and spiral galaxies, respectively.

\begin{table}
\scriptsize
\noindent\begin{minipage}{0.5\textwidth}
\caption{Elliptical Galaxies}
\label{table:elliptical_galaxies}
\begin{tabular}{llccc}
\hline
 & & \scriptsize(Plummer) & \scriptsize(Hernquist) &   \scriptsize(Jaffe) \\
 Galaxy & Type & $m$ (\ev) & $m$ (\ev) & $m$ (\ev)\\
\hline
 NGC  221  & E2 & 184.7 & 199.5 & 212.6\\
 NGC  821  & E4 & 33.8  & 36.7  & 39.4 \\
 NGC 1332  & E6 & 12.7  & 13.3  & 13.7 \\
 NGC 1399  & E1 & 25.8  & 28.6  & 32.4 \\
 NGC 1407  & E0 & 10.2  & 11.1  & 11.8 \\
 NGC 2974  & E4 & 35.7  & 39.7  & 46.1 \\
 NGC 3377  & E5 & 54.2  & 58.9  & 63.2 \\
 NGC 3379  & E1 & 33.6  & 37.3  & 41.9 \\
 NGC 3608  & E2 & 33.8  & 37.5  & 42.6 \\
 NGC 4261  & E  & 33.2  & 35.3  & 44.5 \\
 NGC 4291  & E2 & 43.5  & 48.3  & 55.3 \\
 NGC 4473 & E2 & 31.8  & 35.0  & 38.4 \\
 NGC 4486  & E0 & 22.5  & 24.9  & 28.5 \\
 NGC 4486A & E2 & 43.1  & 46.4  & 49.1 \\
 NGC 4552  & E  & 30.0  & 33.1  & 36.6 \\
 NGC 4564  & E6 & 53.0  & 58.2  & 63.6 \\
 NGC 4621  & E5 & 34.9  & 37.7  & 40.2 \\
 NGC 4649  & E1 & 22.8  & 25.3  & 28.9 \\
 NGC 4697  & E4 & 46.8  & 51.6  & 56.9 \\
 NGC 4742  & E4 & 35.7  & 37.2  & 38.3 \\
 NGC 5077  & E3 & 23.6  & 26.0  & 28.5 \\
 NGC 5576  & E3 & 70.5  & 72.6  & 91.6 \\
 NGC 5813  & E1 & 27.9  & 30.7  & 33.9 \\
 NGC 5845  & E3 & 52.9  & 58.4  & 65.0 \\
 NGC 5846  & E0 & 22.4  & 24.9  & 28.3 \\
 NGC 7052  & E3 & 11.1  & 11.7  & 12.1 \\
 IC4296    & E  & 7.3   & 7.9   & 8.5  \\
 IC1459    & E3 & 19.3  & 21.0  & 22.7 \\
 Cygnus A  & E  & 11.2  & 12.4  & 14.7 \\
 \hline
\end{tabular}

\end{minipage}
\begin{minipage}{0.45\textwidth}
\includegraphics[width=2.3in]{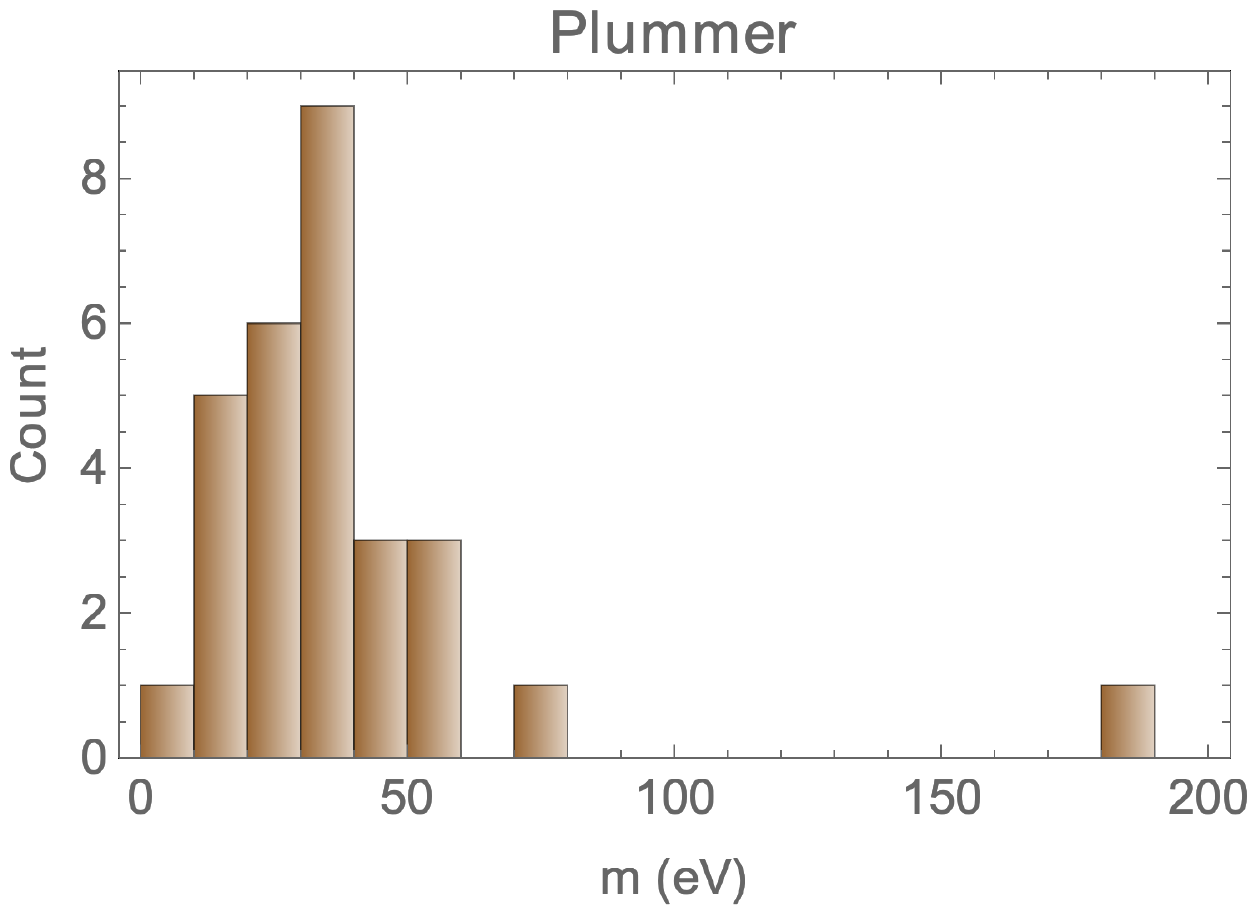}\\
\vspace{20pt}
\includegraphics[width=2.3in]{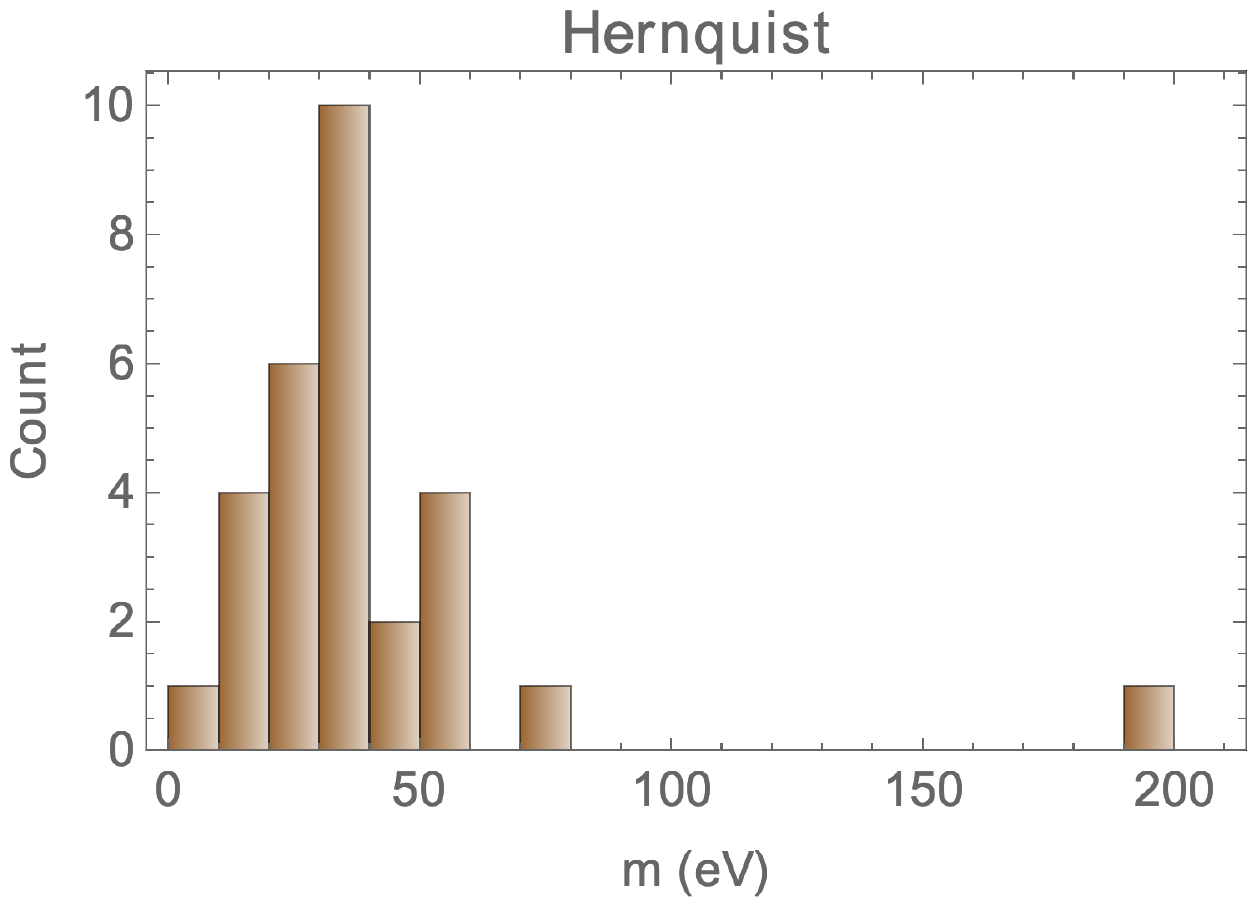}\\
\vspace{20pt}
\includegraphics[width=2.3in]{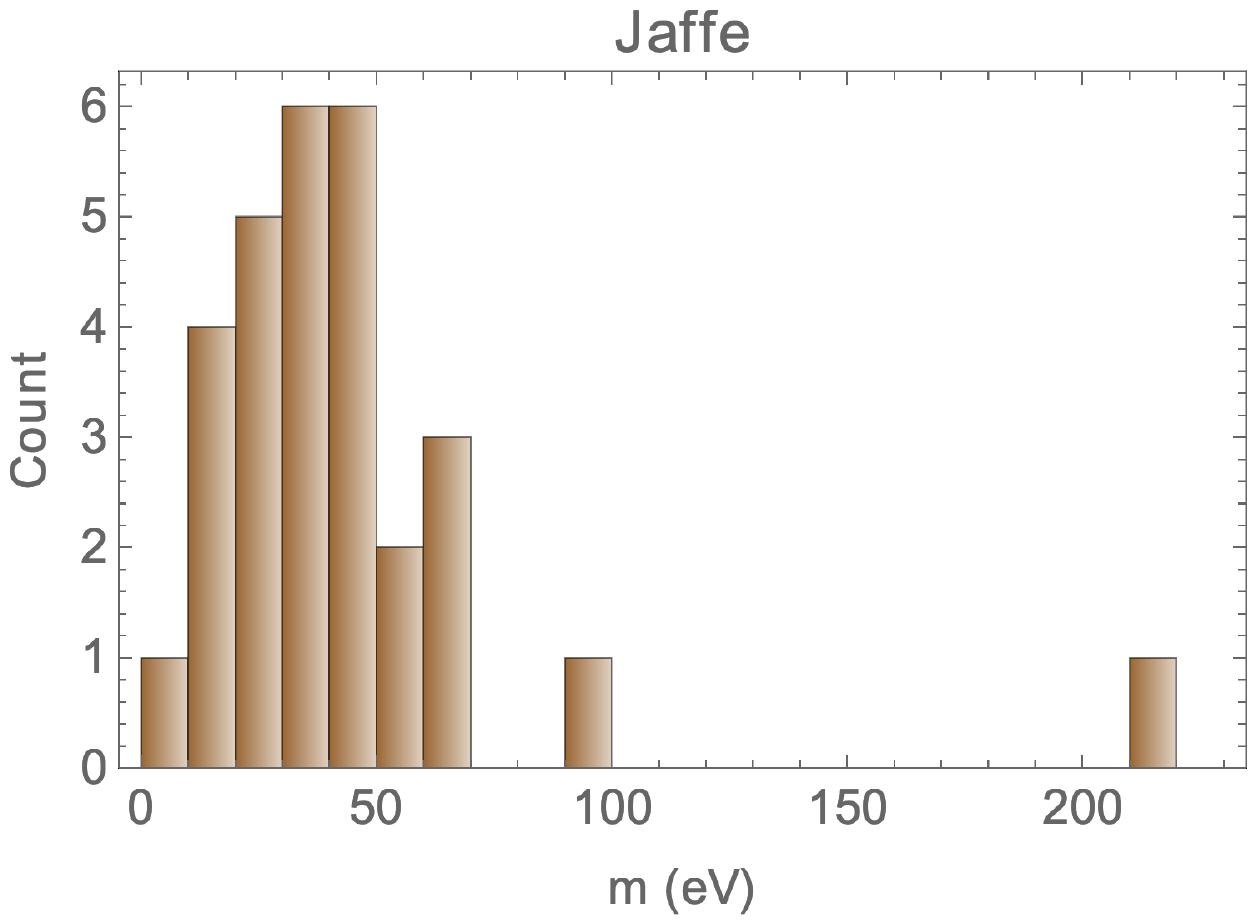}
\end{minipage}
\end{table}

\begin{table}
\scriptsize
\begin{minipage}[18cm]{0.5\textwidth}
\caption{Spiral Galaxies}
\label{table:spiral_galaxies}
\begin{tabular}{llccc}
 \hline
 Galaxy & Type & $m$ (\ev) & $m$ (\ev) & $m$ (\ev)\\
 & & \scriptsize(Plummer) & \scriptsize(Hernquist) & \scriptsize(Jaffe)  \\
 \hline
  Milky Way & Sb & 47.2 & 51.8 & 56.6\\
 NGC 224 & Sb & 55.3 & 60.3 & 65.1\\
 NGC 524 & S0 & 26.4 & 29.3 & 34.0\\
 NGC 1023 & SB0& 40.8  & 43.5 &  45.6\\
  NGC 1068 & SBb & 32.6 & 35.2 & 37.4\\
 NGC 1316 & SB0 & 20.4 & 21.7 & 22.7\\
  NGC 1227 & S0 & 26.7 & 28.7 & 30.8\\
 NGC 2549 & S0 & 62.3 & 69.1 & 77.7\\
  NGC 2787 & SB0 & 91.8 & 101.4 & 112.6\\
 NGC 3031 & Sab & 46.0 & 49.8 & 53.2\\
 NGC 3115 & S0 & 51.9 & 55.7 & 58.9\\
 NGC 3227 & SBa & 86.5 & 148.1 & 192.0\\
  NGC 3384 & SB0 & 30.1 & 31.2 & 31.9\\
 NGC 3245 & S0 & 42.3 & 46.5 & 50.9\\
 NGC 3414 & S0 & 38.1 & 41.6 & 45.0 \\
 NGC 3585 & S0 & 20.4 & 22.0 & 23.4\\
 NGC 3607 & S0 & 28.2 & 31.3 & 35.3\\
  NGC 3945 & SB0 & 78.3 & 66.5 & 89.9\\
 NGC 3998 & S0 & 38.4 & 40.9 & 46.1\\
 NGC 4026 & S0 & 52.5 & 58.2 & 67.2\\
 NGC 4151 & Sa & 40.6 & 43.7 & 46.3\\
 NGC 4258 & SBbc & 79.3 & 87.8 & 98.3\\
 NGC 4459 & S0 & 38.9 & 41.7 & 44.1\\
 NGC 4596 & SB0 & 45.7 & 49.8 & 53.9\\
 NGC 5128 & S0 & 45.0 & 48.5 & 51.6\\
 NGC 7457 & S0 & 91.0 & 101.2 & 118.3\\
 NGC 3079 & SBcd & 40.8 & 43.6 & 45.8\\
 NGC 3393 & Sba & 34.2 & 37.9 & 42.4\\
 Circinus & Sb & 62.4 & 66.1 & 69.0\\
 IC2560 & SBb & 34.1 & 36.4 & 38.4\\
 P49940 & S0 & 136.6 & 156.1 & 122.2\\
  \hline
\end{tabular}

\end{minipage}
\begin{minipage}{0.45\textwidth}
\includegraphics[width=2.3in]{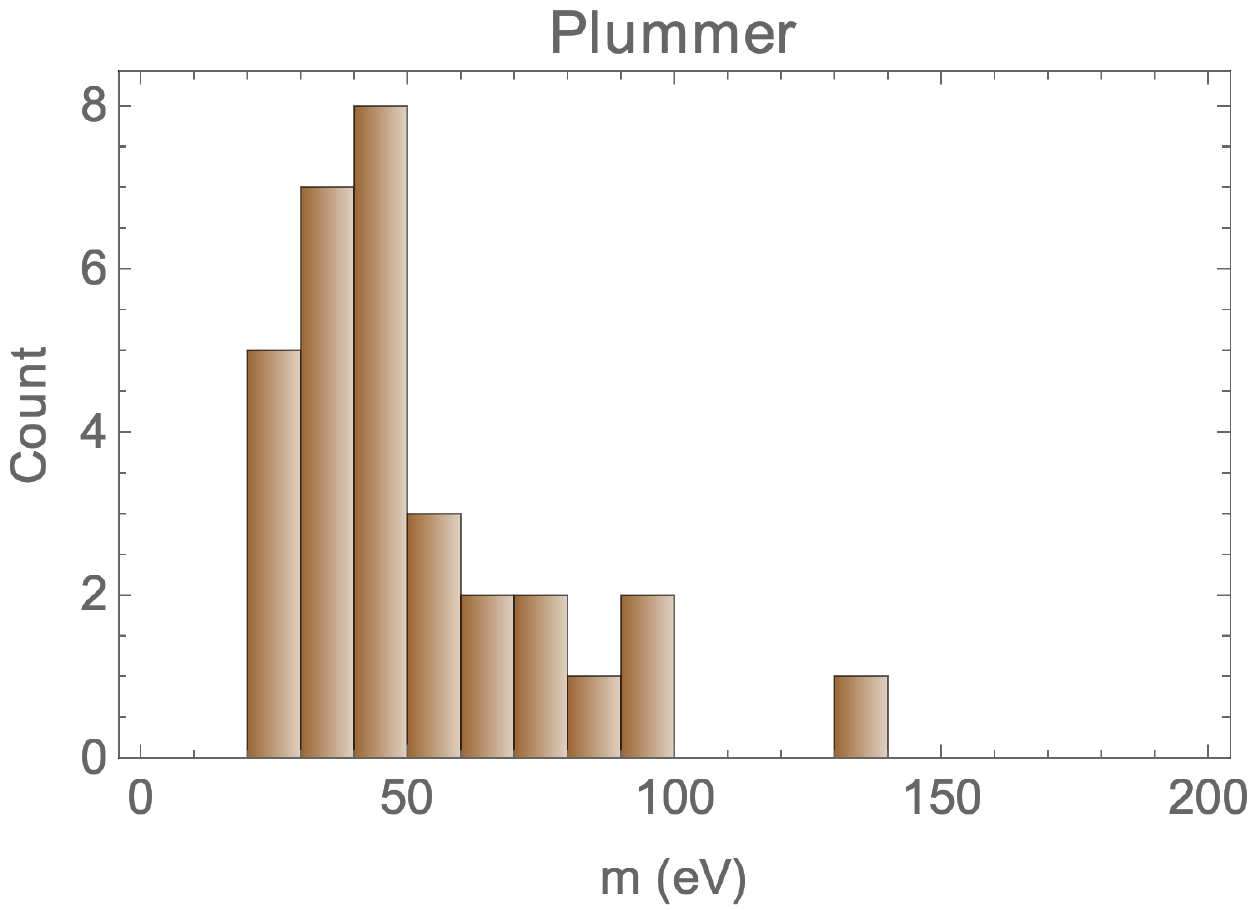}\\
\vspace{20pt}
\includegraphics[width=2.3in]{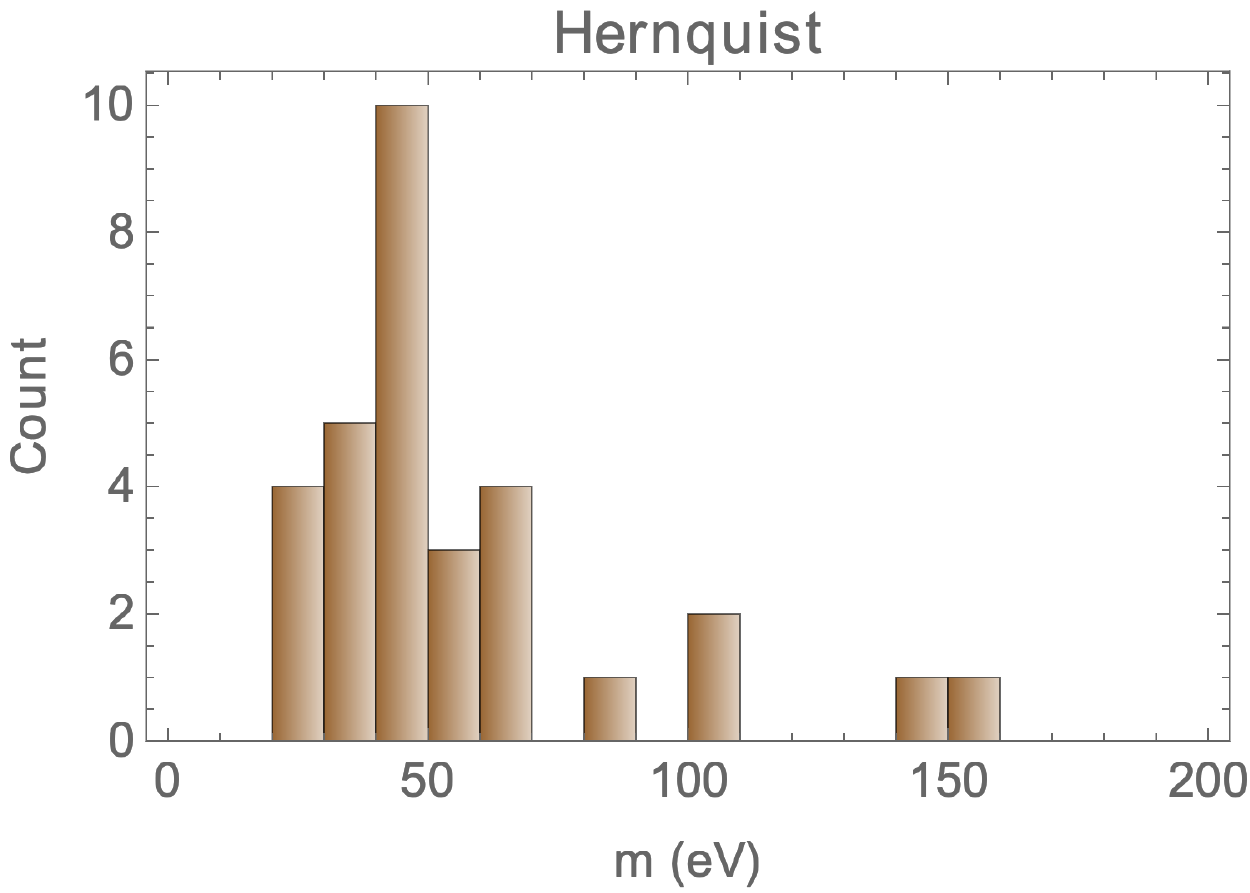}\\
\vspace{20pt}
\includegraphics[width=2.3in]{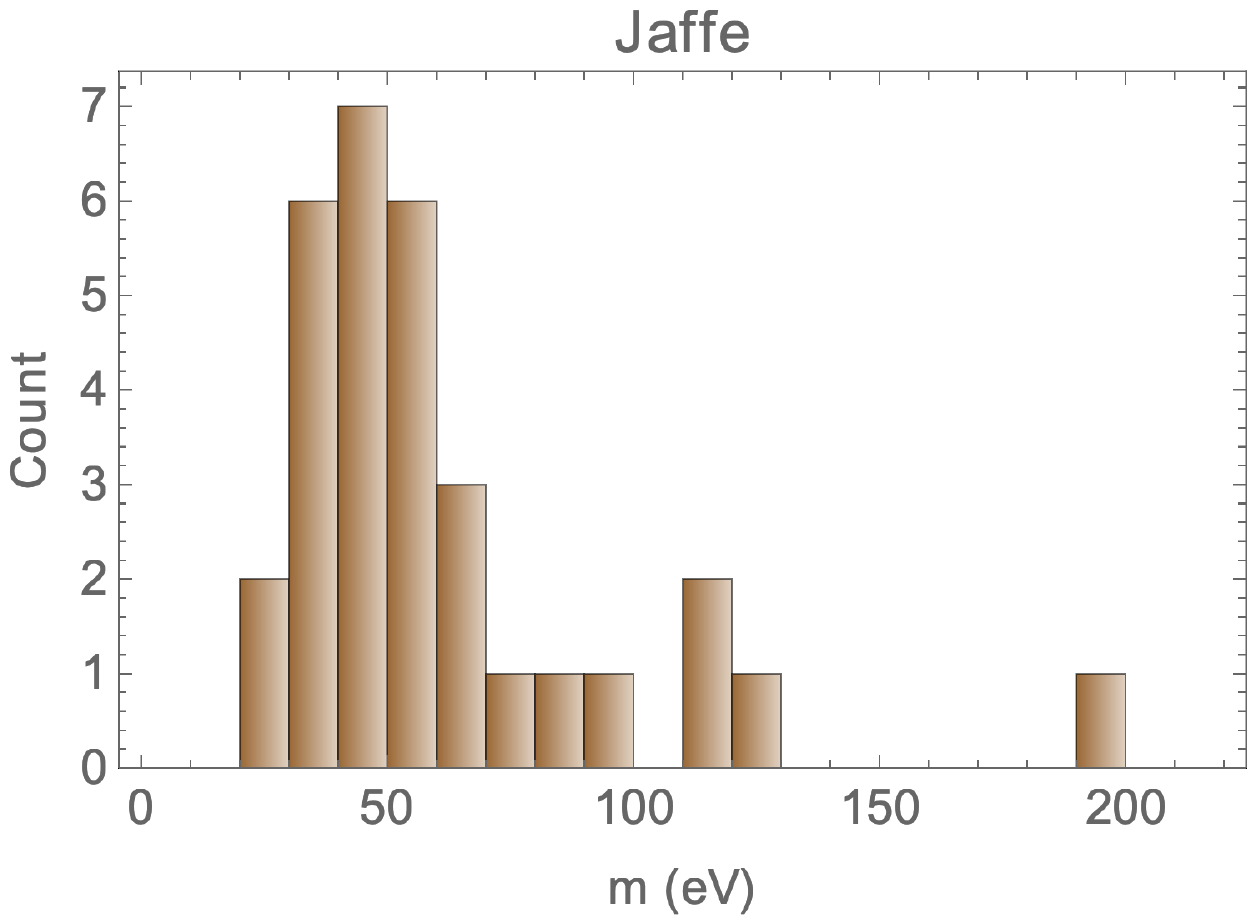}
\end{minipage}
\end{table}

For spiral galaxies, we find that the DM mass lies in the range $30-100$ eV with a few outliers in the range $ \sim 100-150$ eV. For elliptical galaxies, $m$ has a tighter range, $10-60$ eV for all the three baryon profiles (excluding the one outlier, NGC 221). The average and standard deviation for the calculated DM mass for the two different galaxy types and three baryonic profiles are listed in  table \ref{mean_std}.

\begin{table}[!htb]
\caption{Statistics of the DM Mass}
\label{mean_std}
\vspace{-15pt}
$$
\begin{array}{|c|cc|cc|} 
 \hline
  &  \multicolumn{2}{|c|}{\rm Elliptical \ Galaxies } &  \multicolumn{2}{|c|}{\rm Spiral \ Galaxies }  \cr
 \hline
\rm  Baryon Profile &  \rm avg (eV)  &\rm  std. dev. (eV) & \rm avg (eV) & \rm std. dev. (eV) \cr
 \hline
\rm Plummer & 36.8 & 32.1 & 50.6 & 25.6\cr
\rm Hernquist & 40.0 & 34.6 & 56.5 & 32.4 \cr
\rm  Jaffe & 44.5 & 37.4 & 61.7 & 36.0 \cr
\hline
\end{array}
$$
\end{table}

It is important to note that the average value of $m$ for elliptical galaxies is lower than that for spiral galaxies. This is due, to a great extent, to having ignored the spiral mass in the above calculations: if we add the spiral mass to the bulge  and increase the effective radius (while keeping all other parameters fixed), the value of $m$ decreases considerably. For example, in the Milky Way (spiral mass  $5.17 \times 10^{10} M_{\odot} $, bulge mass $0.91 \times 10^{10} M_\odot$  \cite{licquia2015improved}), this shifts the DM mass from $51.8$ eV to $22.38$ eV  for a change in effective radius from 0.7 kpc to 3 kpc (for Hernquist profile). Even though adding the entire baryonic mass of the spiral to the bulge stellar mass by  just increasing the bulge effective radius is probably a poor assumption, it can be expected that considering the disc structure would lead to a decrease in the mean value of $m$, closer to the result for elliptical galaxies. On the other hand $\vrotb$ is not known for most bulge-dominated elliptical galaxies, so uncertainties in this parameter may shift the DM mass for ellipticals, but the change in that would be less significant. Overall, it is remarkable that despite all its simplifying assumptions the model provides values of $m$ that lie within a relatively narrow range~\footnote{The case of fermionic DM for the Milky way considering most of the structural features of the galaxy has been studied \cite{barranco2018constraining}. However, they assume complete degeneracy at zero temperature and the mass range is obtained strictly from the constraints on the rotation curve.}. 

\begin{figure}[ht]

$$
\includegraphics[width=2.3in]{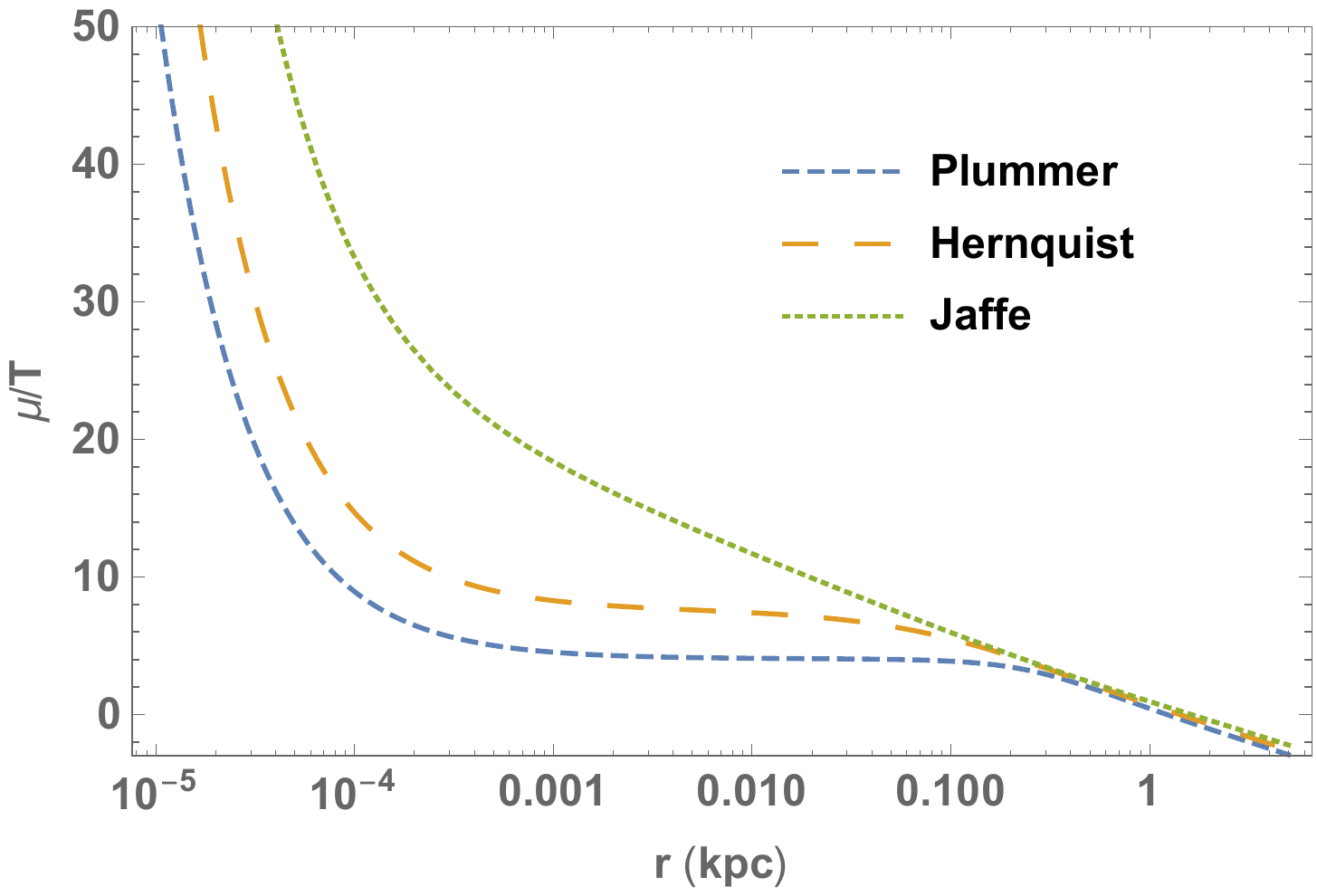}
\includegraphics[width=2.3in]{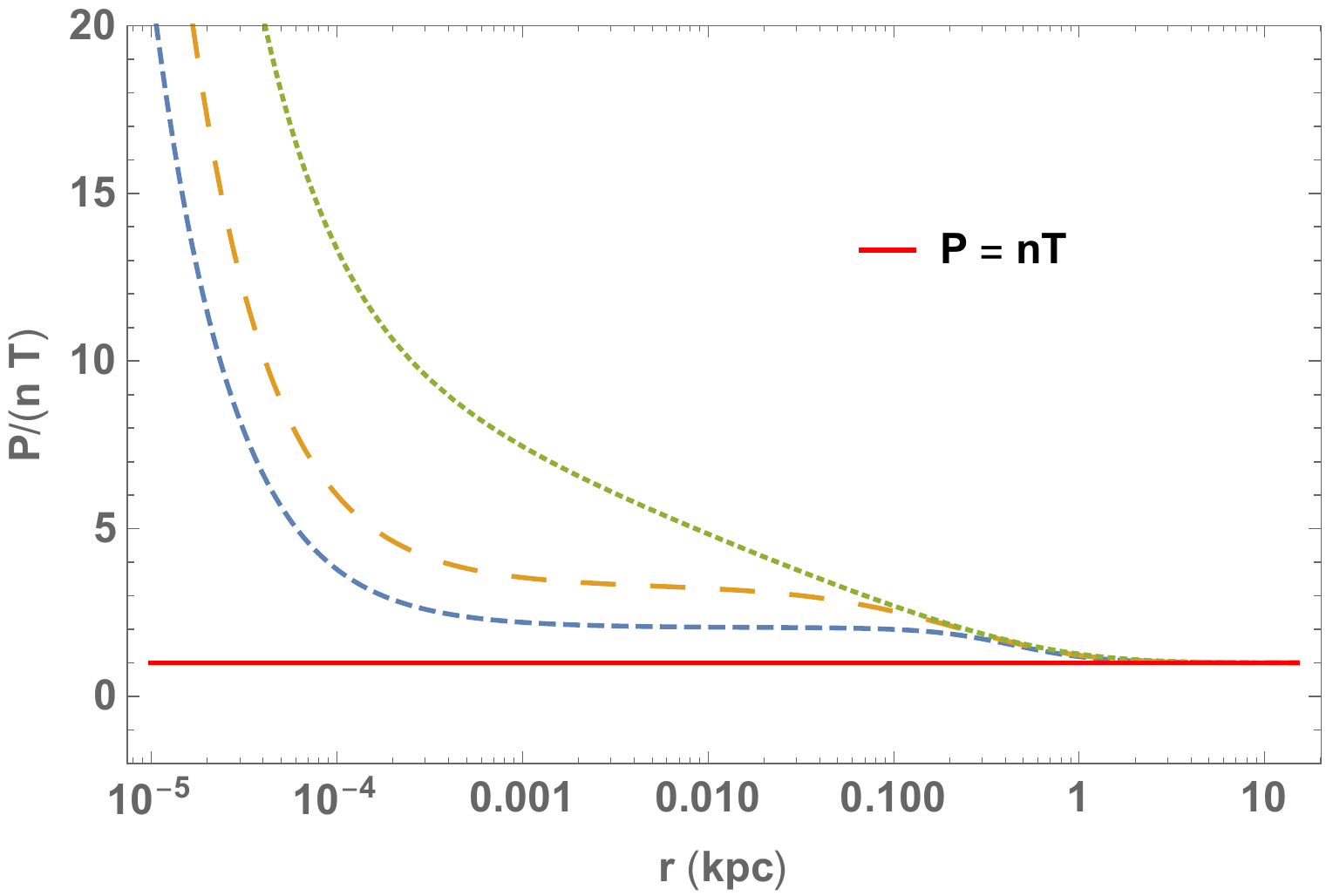} 
\includegraphics[width=2.4in]{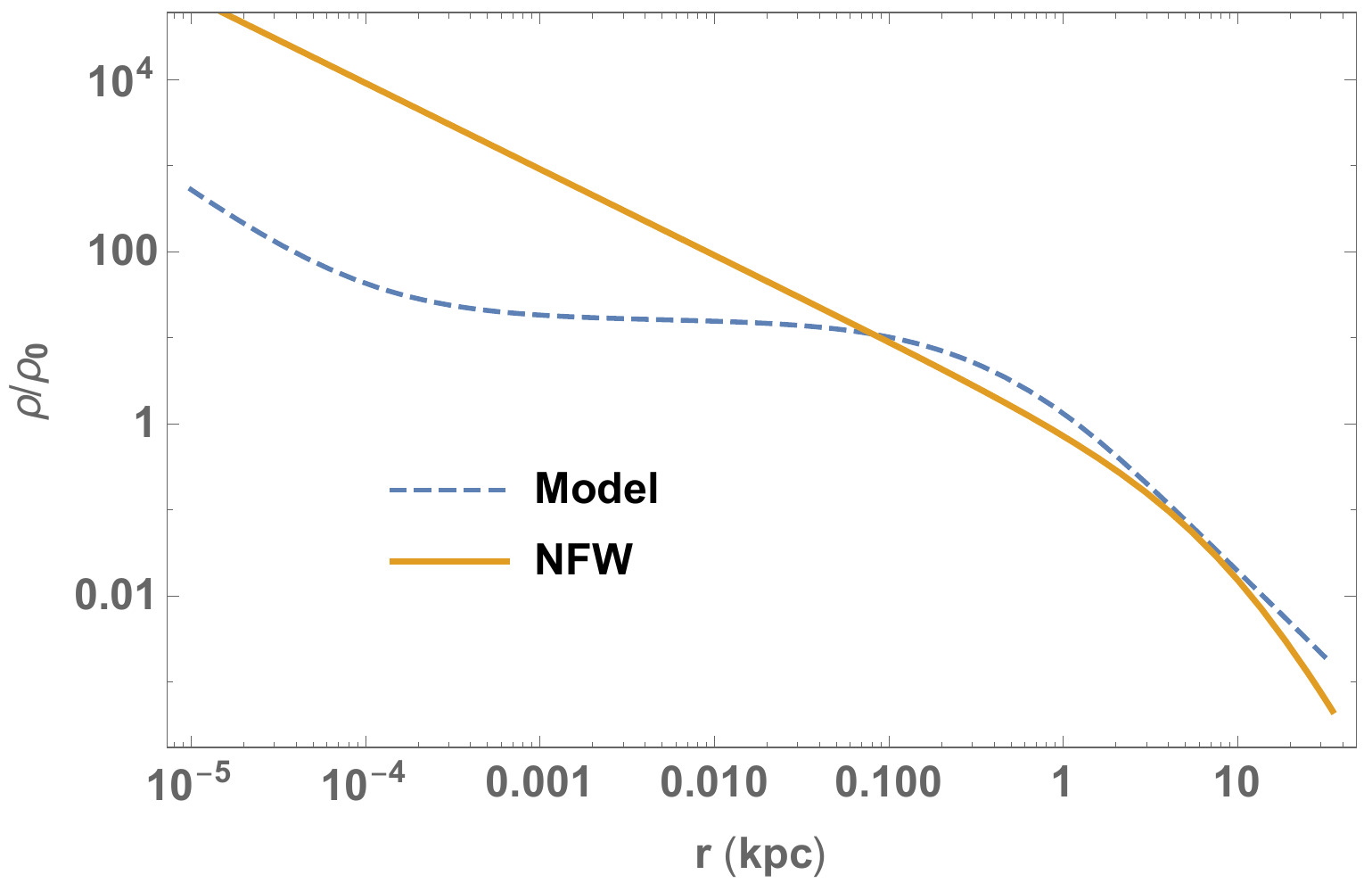}
$$
	\caption{\footnotesize DM chemical potential (left) and $P/(n T)$ (middle) as functions of $r$  for the Milky Way  (see  \cref{eq: n,eq:sigma,eq:T-vrot}) for three baryon density profiles; the classical Maxwell-Boltzmann equation of state is shown in red. Right: comparison of the DM density profile for the model discussed here using the Hernquist profile with the (unnormalized) NFW profile \cite{sofue2012grand} to illustrate the presence of a core in the former. All graphs are for the Milky Way. }
\label{fig:mu-PnT}
\end{figure}

The histograms next to tables \ref{table:elliptical_galaxies} and \ref{table:spiral_galaxies} exhibit a few ``outliers'', for which the DM mass is in the $ \gtrsim 100 $ eV range, though this is dependent on the baryon profile used. For example, $m$ associated with NGC 2778 is $ \sim 75$ eV for the Plummer and Hernquist profiles, but $ \sim 100$ eV for the Jaffe profile, while $m$ for NGC 6068 and NGC 5576 exhibit the opposite behavior. The case of NGC 221 is unique in that it requires $ m \sim 200 $ eV, but it is also special in that it is the smallest galaxy in this set (with an effective radius of $40$ pc), and categorized as a dwarf galaxy with a central black hole.  By comparing the bulge mass from two different sources (log($\mb$)  of 9.05 in \cite{kormendy2013coevolution} and 8.53 in \cite{hu2009black}) hint at larger uncertainties in the measurement of stellar mass and lead to a comparatively large value for the DM mass. 

\begin{figure}
\bal
\includegraphics[width=2.3in]{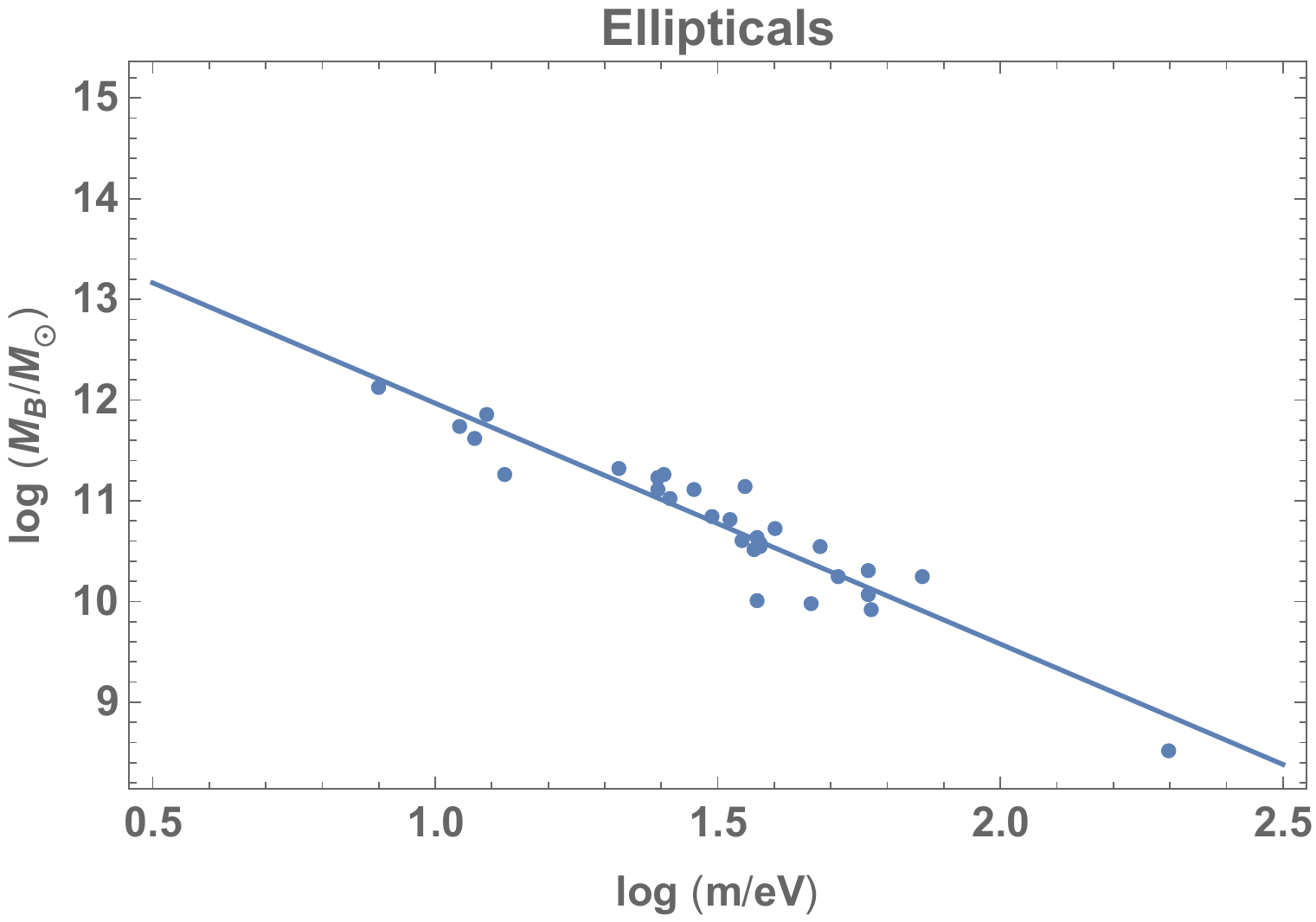} \qquad & \qquad \includegraphics[width=2.3in]{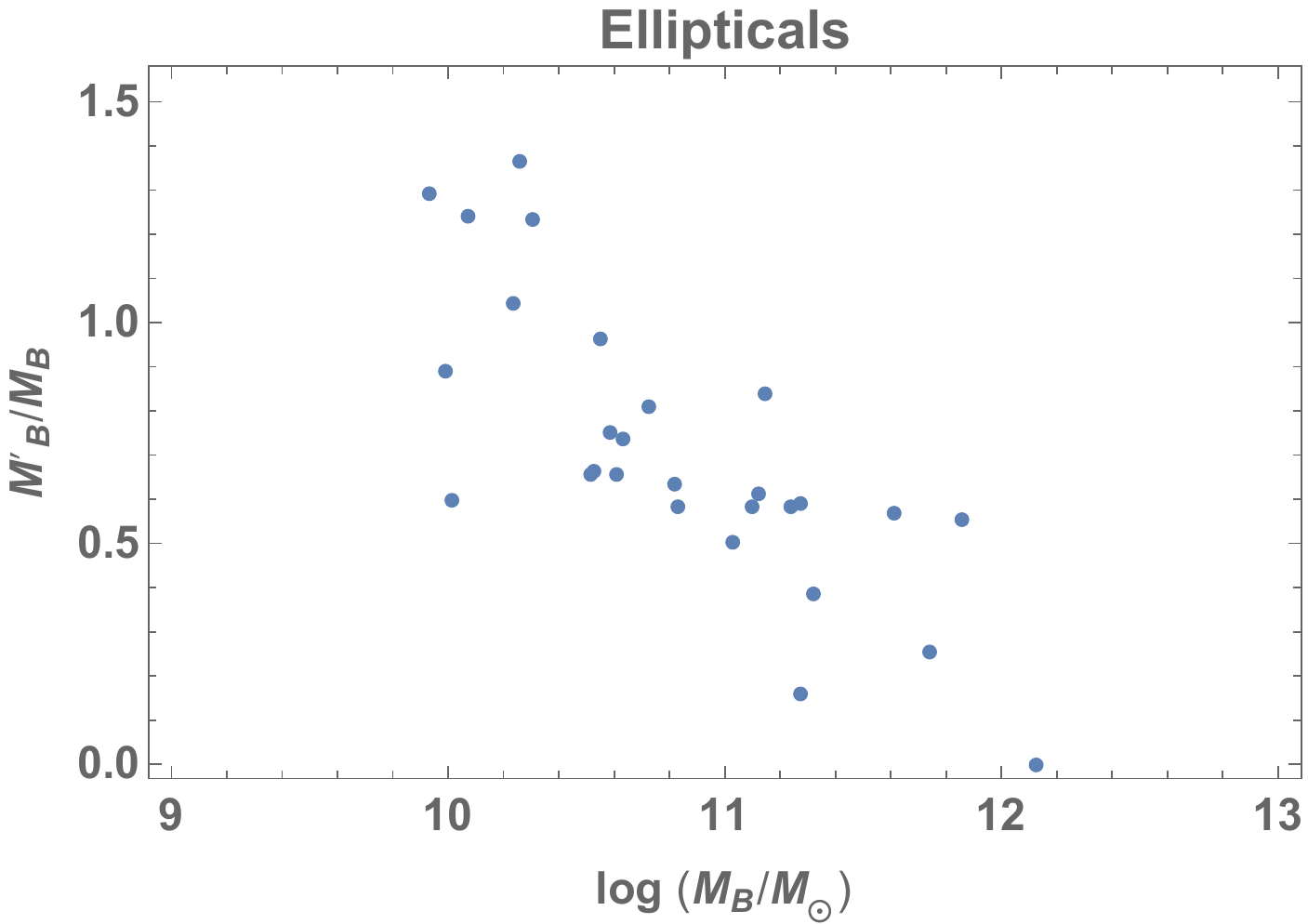}
 \mcr
\includegraphics[width=2.3in]{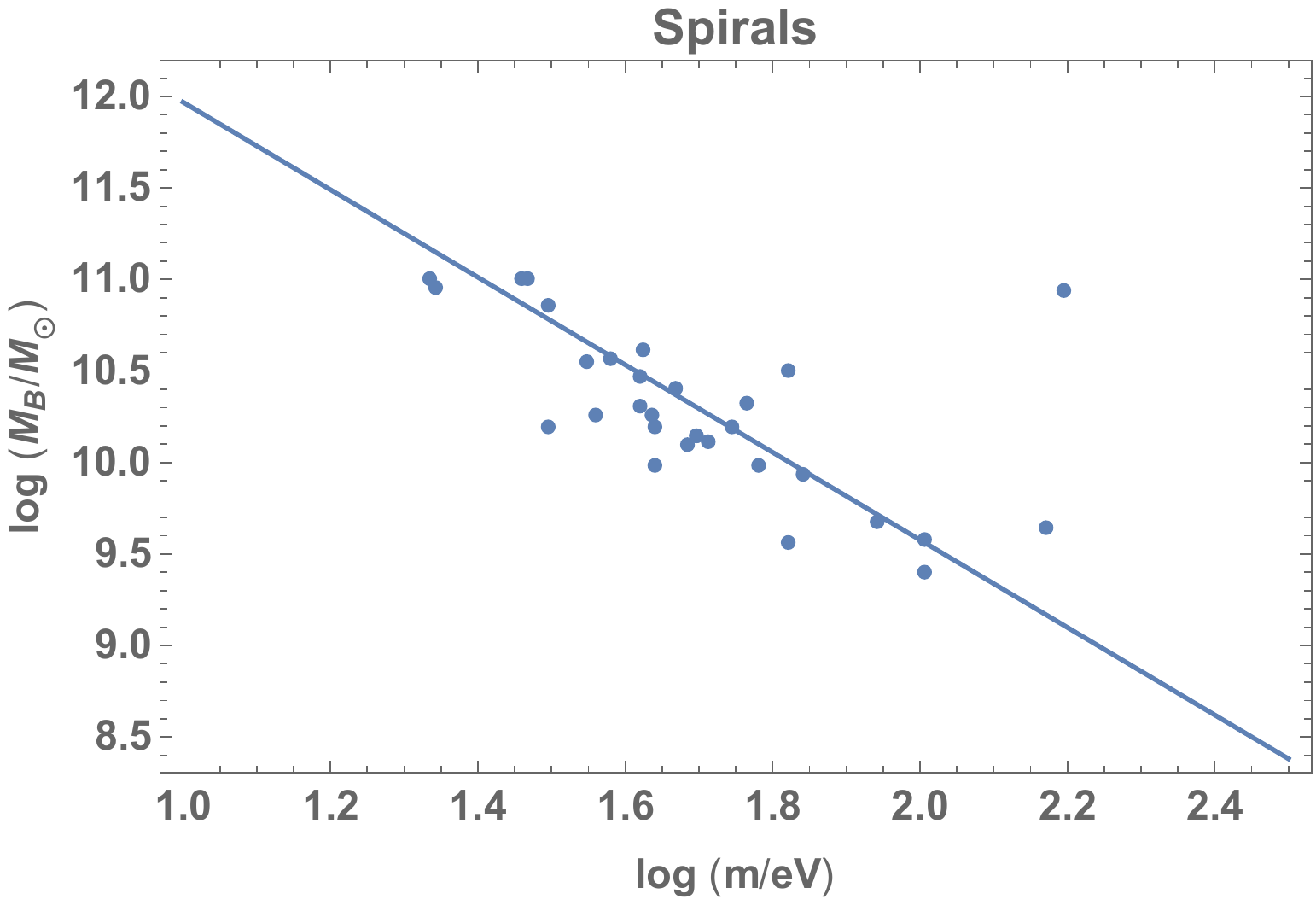} \qquad & \qquad\includegraphics[width=2.3in]{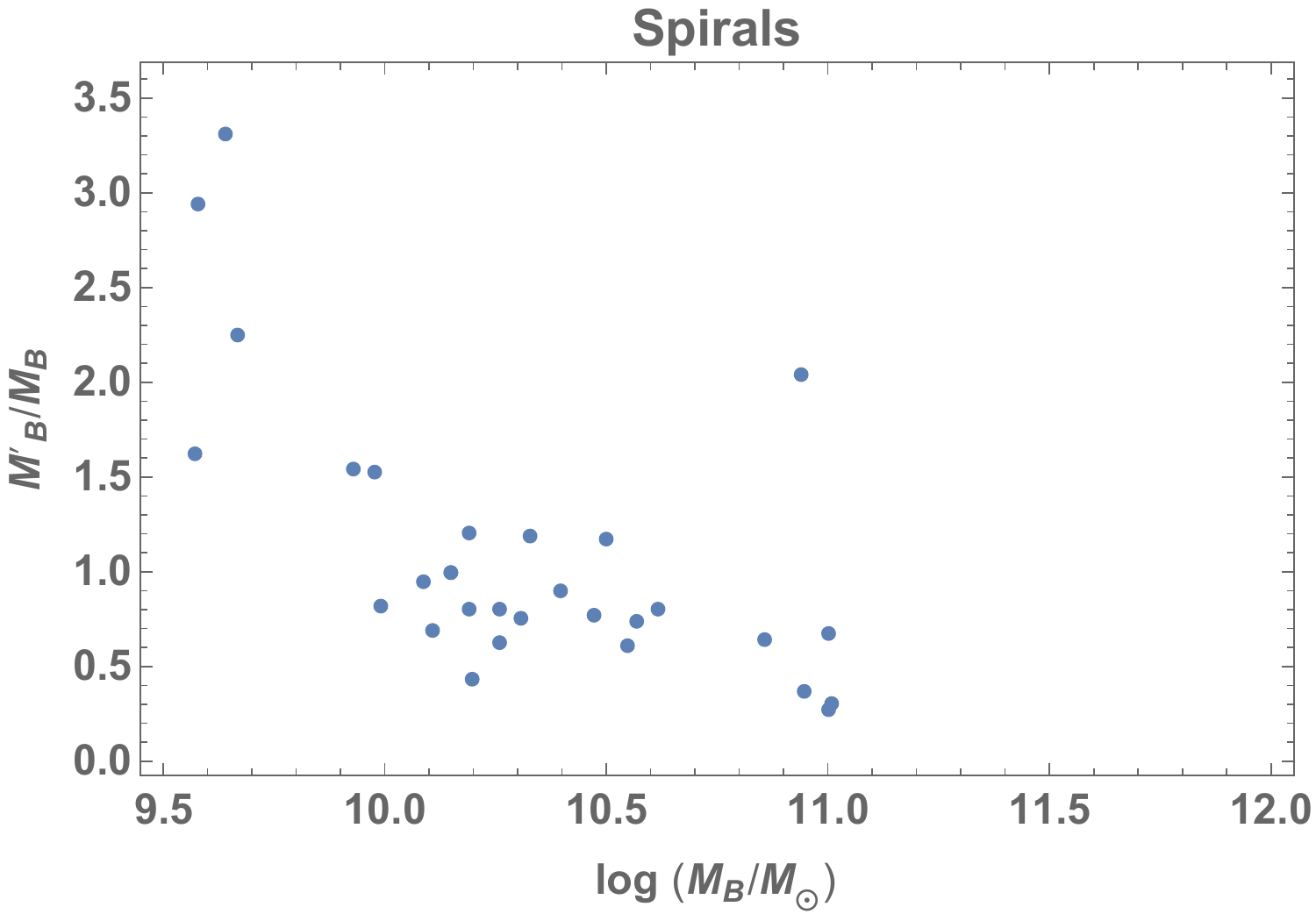} 
\nonumber \end{align}
\caption{Left: scatter plot and linear fit illustrating the correlation between the obtained values of $m$ and $\mb$ for elliptical (top) and spiral (bottom) galaxies. Right: relative shift in $\mb$ needed to obtain a fixed value of $m$, chosen here as $50$ eV,  for elliptical (top) and spiral (bottom) galaxies. NGC 221 is not included in the plots. All the results are for the Hernquist profile.}
\label{fig:dMB-MB}
\end{figure}

To further understand the spread of $m$ values we present in Fig. \ref{fig:dMB-MB} a plot of $m$ against $\mb$  for the galaxies in our dataset, where we find that larger values of $m$ are  associated with smaller, less massive galaxies. This correlation may indicate a defect in the DM model (which should produce similar values of $m$ for all galaxies, without the correlation show in the figure), or it may indicate that the data we use  underestimates $\mb$ for smaller galaxies, and over-estimates it for larger ones. To examine this last possibility we took from our dataset the values of $ \vrotb$ and $ \rb$ for each galaxy and then obtained the baryon mass that corresponds to a fixed choice of $m=50 $eV. We denote this `derived' baryon mass by $ \mb'$,  In Fig. \ref{fig:dMB-MB} we also present a plot of $ M'_B / \mb $ vs  $\mb$, which shows that $ |  M'_B| \lesssim 3 \mb$ for the spiral galaxies in our set, and  $ |  M'_B| \lesssim  1.5 \mb$ for the ellipticals, so that an $\mathcal{O}(1)$ shift in $\log \mb$  can explain the fact that we do not obtain the same value of $m$ for these galaxies.  Although we believe this argument is compelling, factors of order $\sim 2$-$3$ can easily be accommodated given the current systematic  errors in the estimation of $\mb$ associated to stellar evolution, reddening and the past star formation history of each galaxy (see for instance \cite{Bell2001}). Therefore the viability of the dark matter model in this context then cannot be absolutely decided.

We now consider various aspects of the solutions to \cref{eq:eom}, using the Milky Way as an example. In Fig. \ref{fig:mu-PnT}, we show the chemical potential for three different baryon profiles. As expected, $\mu(r)$ diverges as $r$ approaches the galactic center, indicating  a the presence of a SMBH. We also examine the degree to which the gas is degenerate by plotting $ P/(n T) $. Far from the galactic center, the gas obeys the classic (dilute) Maxwell-Boltzmann distribution $ P \simeq n T$ (red line in the figure), while close to the galactic center, a significant deviation due to Fermi-Dirac statistics is observed, indicating strong degeneracy. In the bottom panel of Fig.~\ref{fig:mu-PnT} we compare the obtained density profile in the inner regions to the empirical solution found for collisional cold dark matter model, or NFW profile \citep{navarro1996structure}. At the centers of halos, the cold dark matter solution is characterized by a cuspy mass distribution while our model favors shallower inner dark matter cores, with the exception of the region surrounding the central black hole.

\begin{figure}
\bal
\includegraphics[width=2.3in]{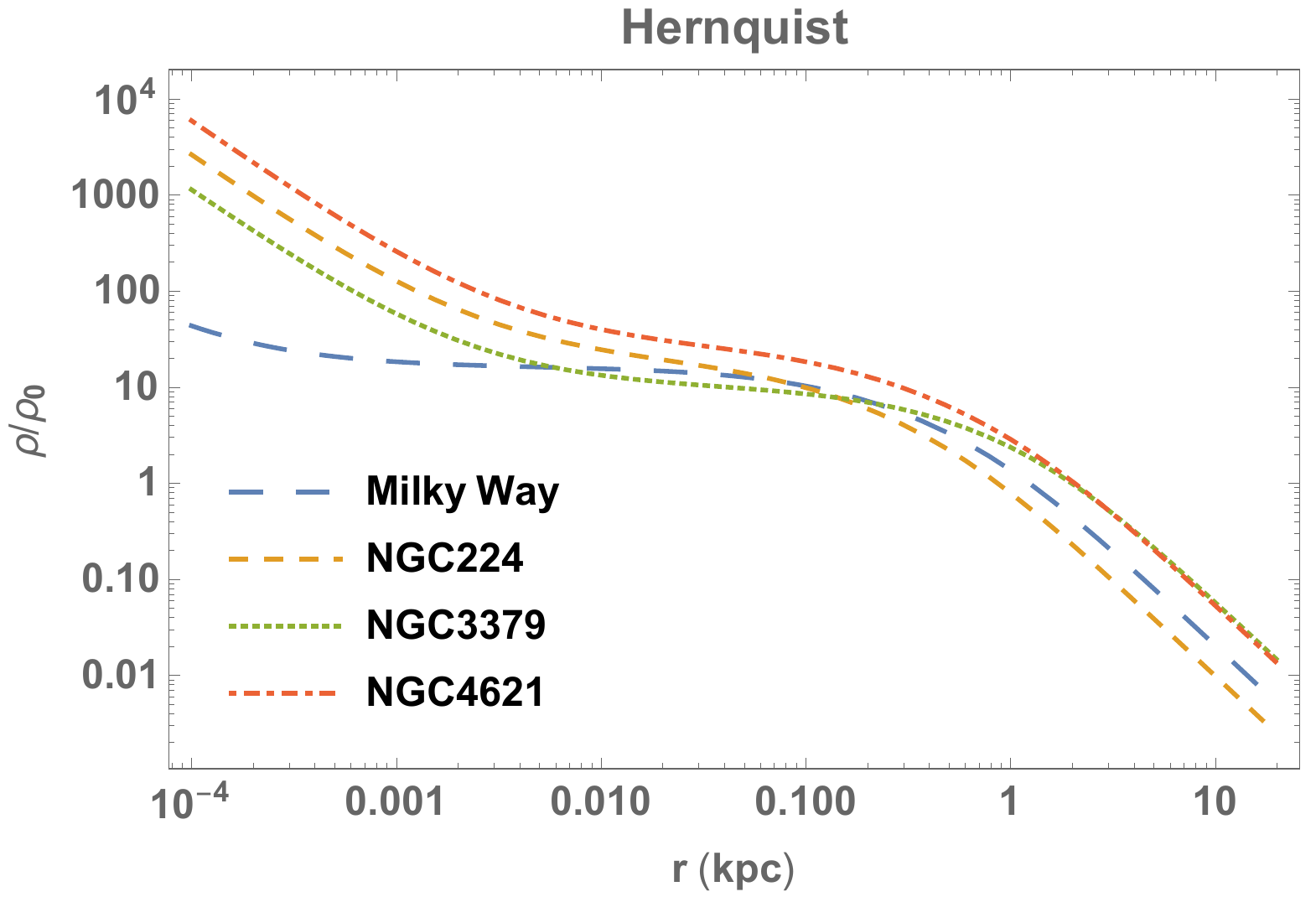} \qquad & \qquad \includegraphics[width=2.3in]{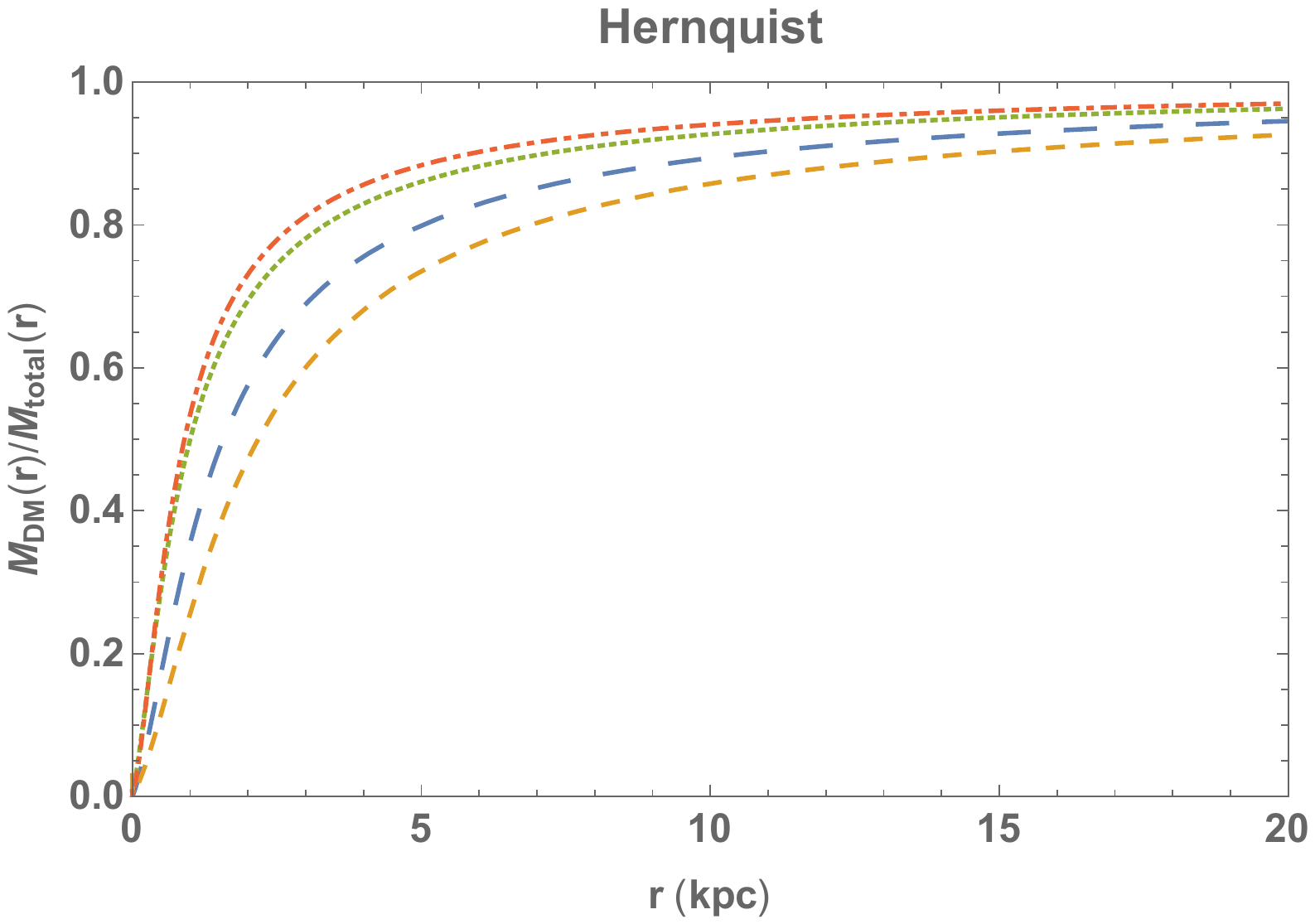} \mcr
\includegraphics[width=2.3in]{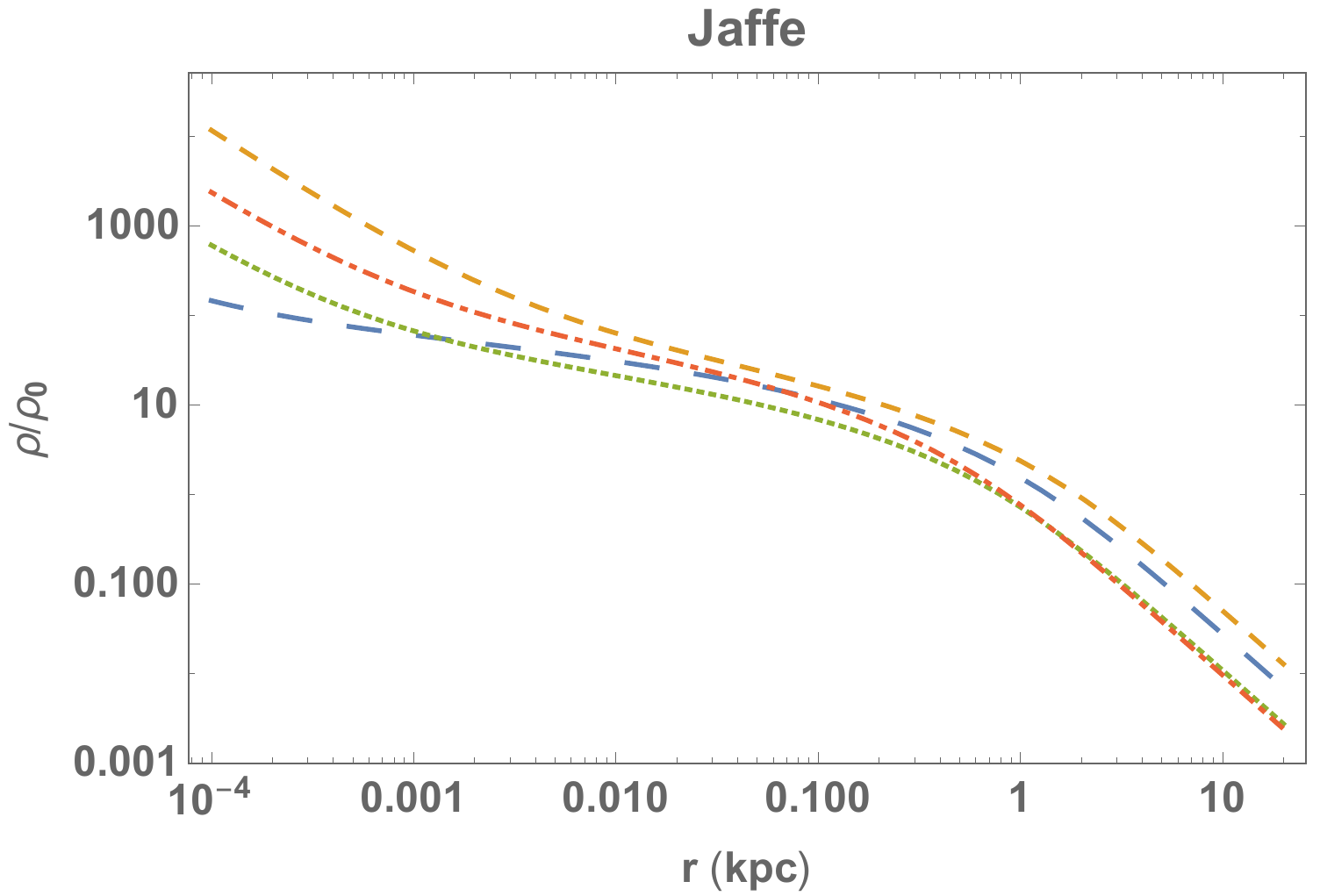} \qquad & \qquad\includegraphics[width=2.3in]{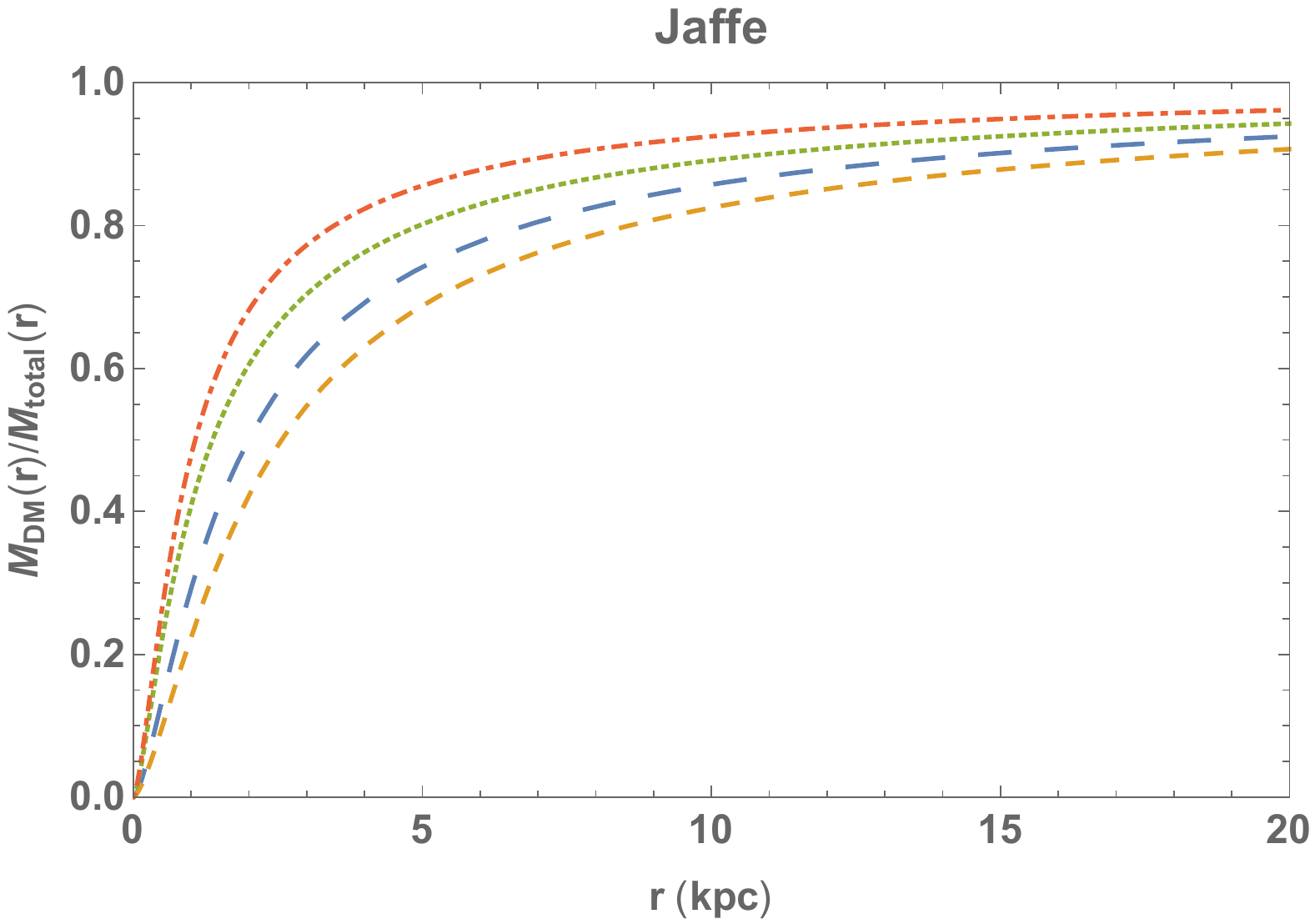} \mcr
\includegraphics[width=2.3in]{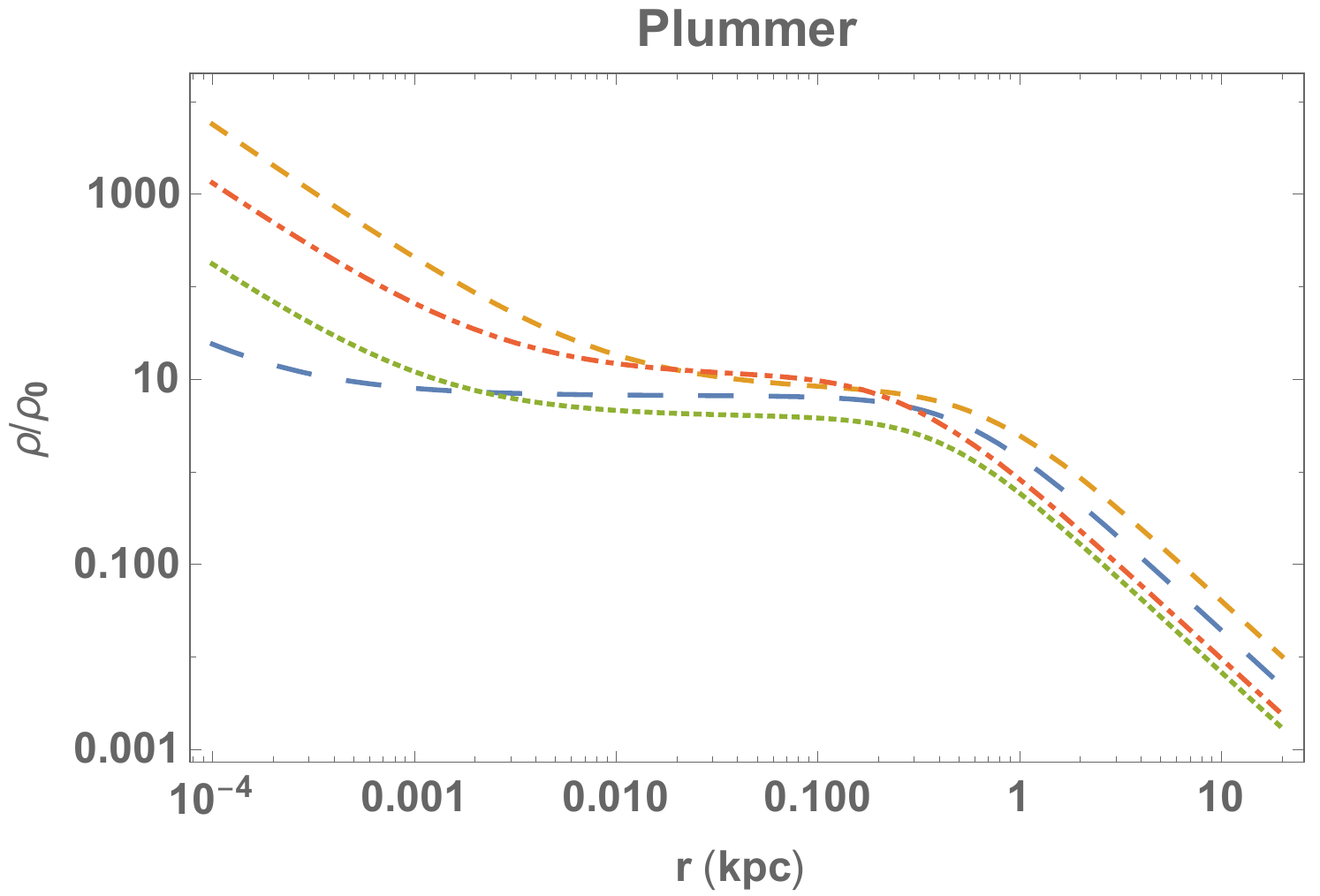} \qquad & \qquad \includegraphics[width=2.3in]{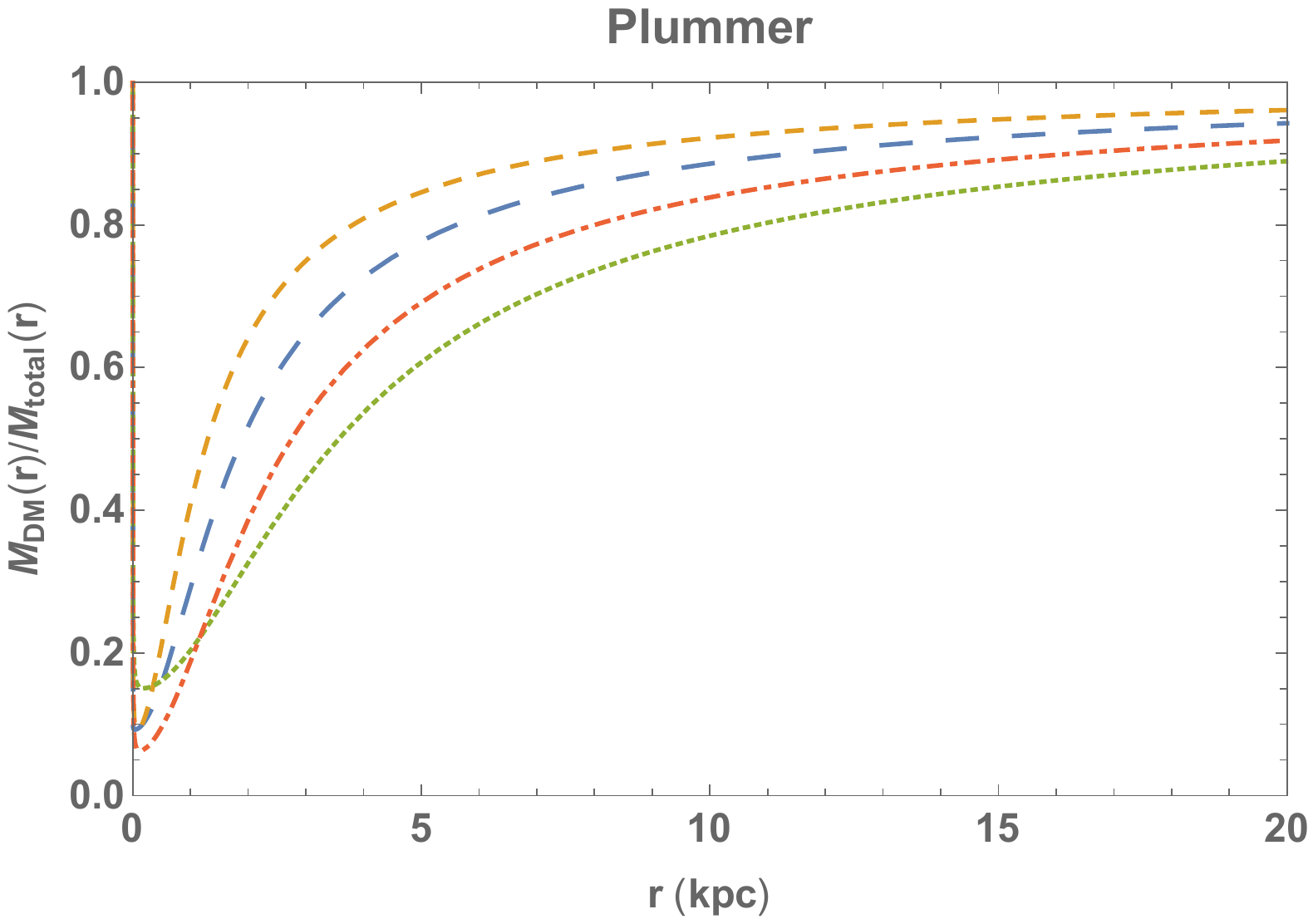}
\nonumber \end{align}
\caption{DM density (left column) and mass fraction (right column) for two spiral galaxies (Milky Way and N224) and two elliptical galaxies (N3379 and N4621) (we use $ \rho_0 = 2 m/\lambda^3$). }
\label{fig:density-mf}
\end{figure}

\begin{figure}[ht]
\bal
\includegraphics[width=2.4in]{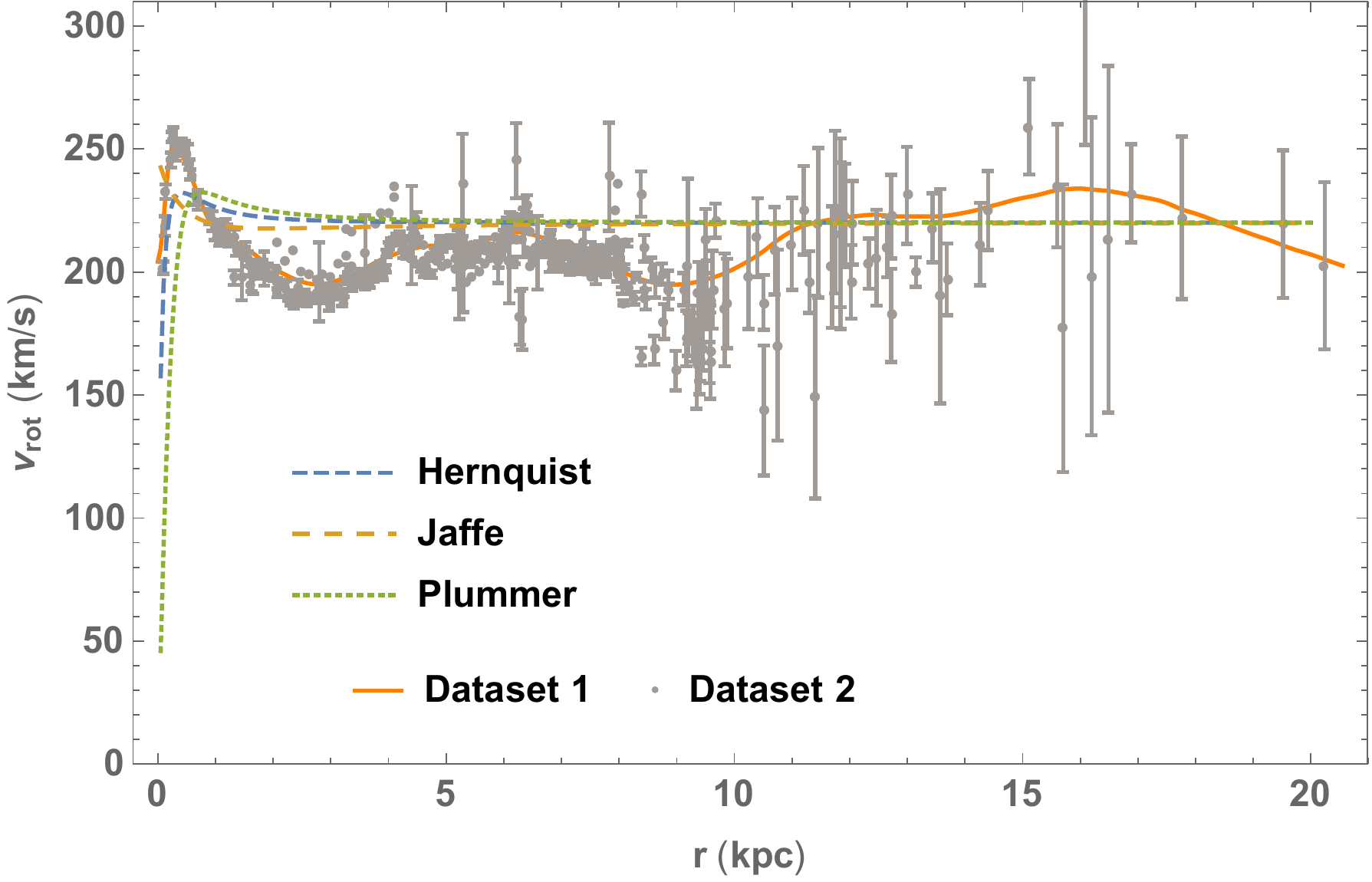} \qquad & \qquad \includegraphics[width=2.4in]{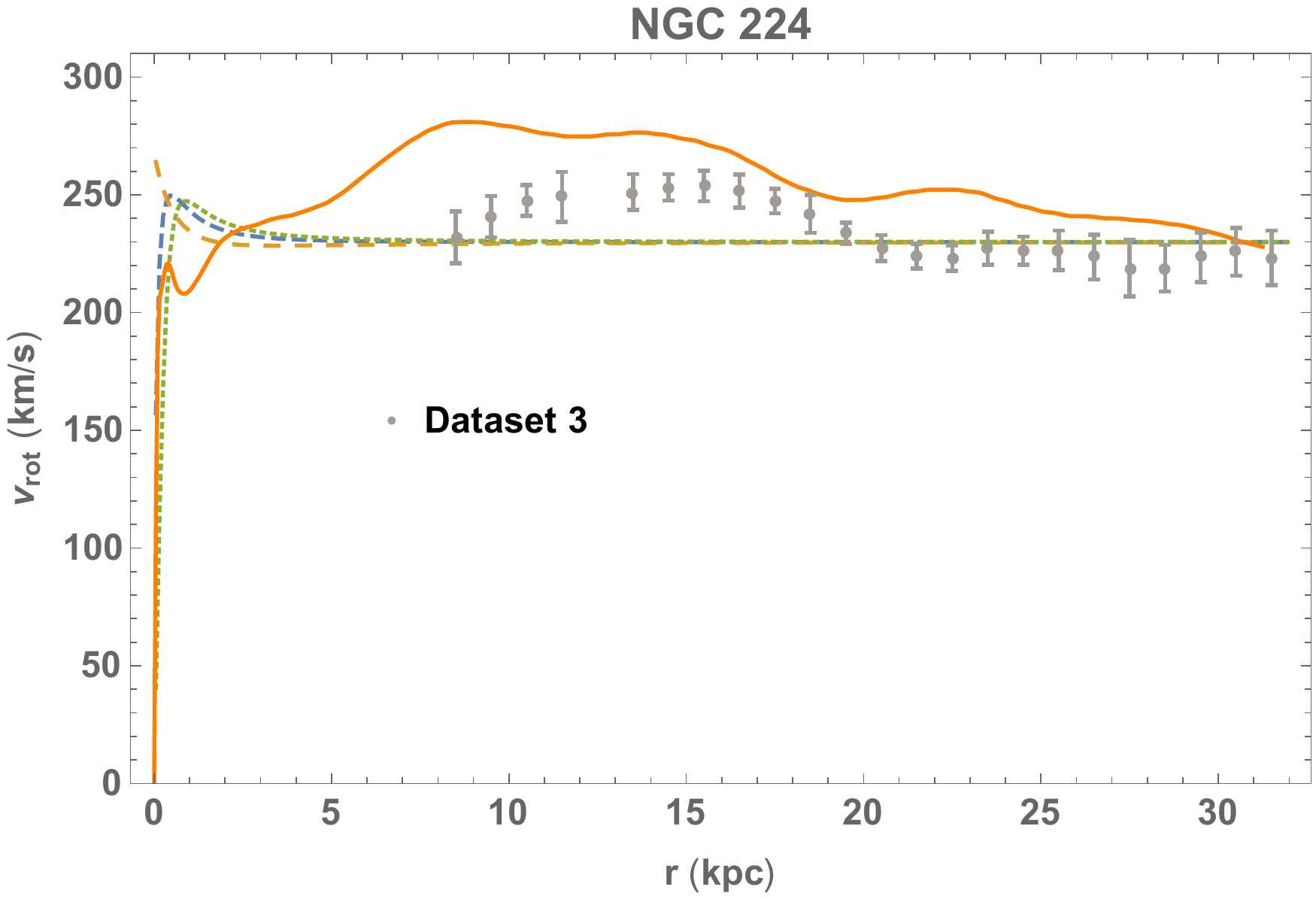}\mcr
\includegraphics[width=2.4in]{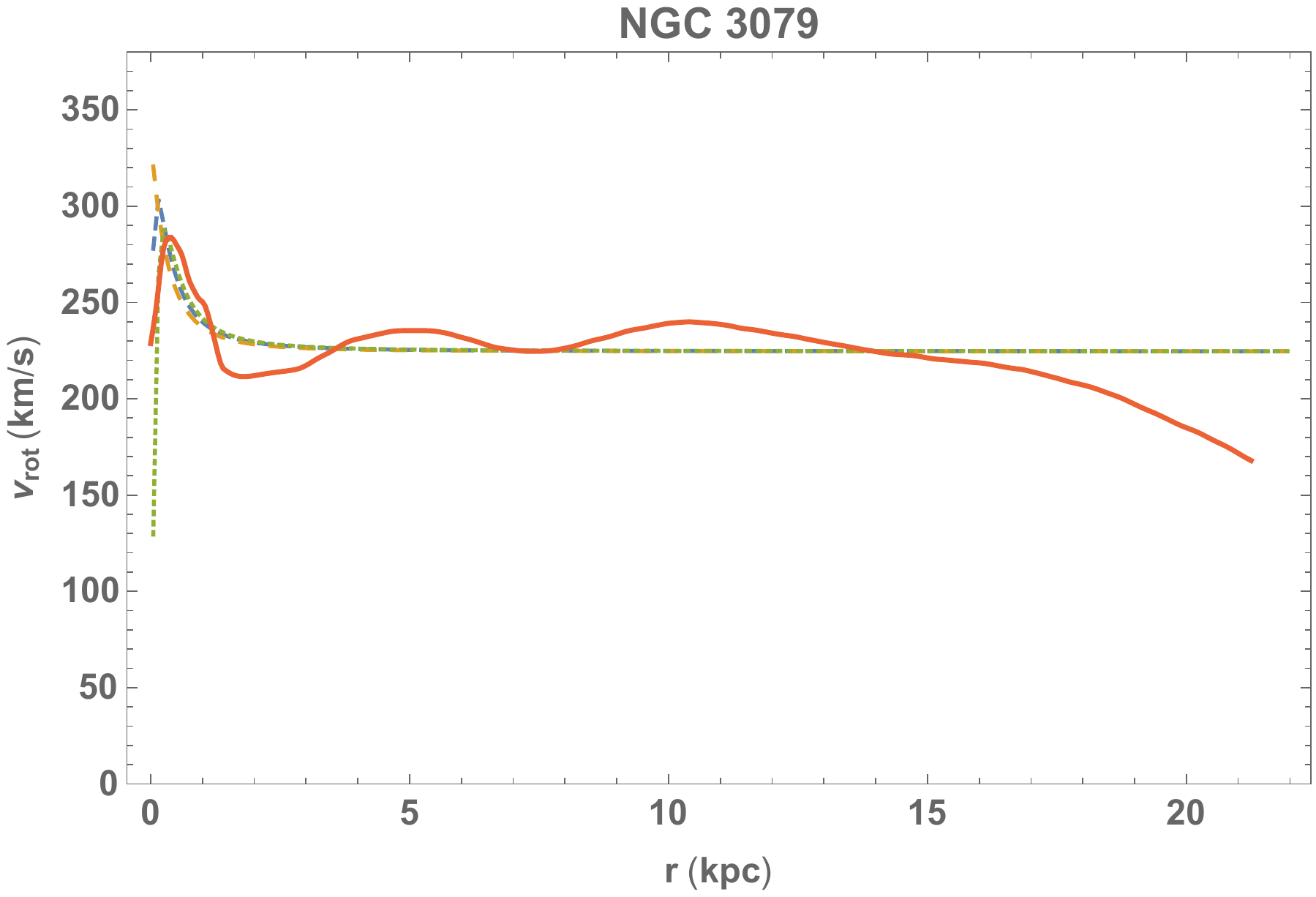}  \qquad & \qquad \includegraphics[width=2.4in]{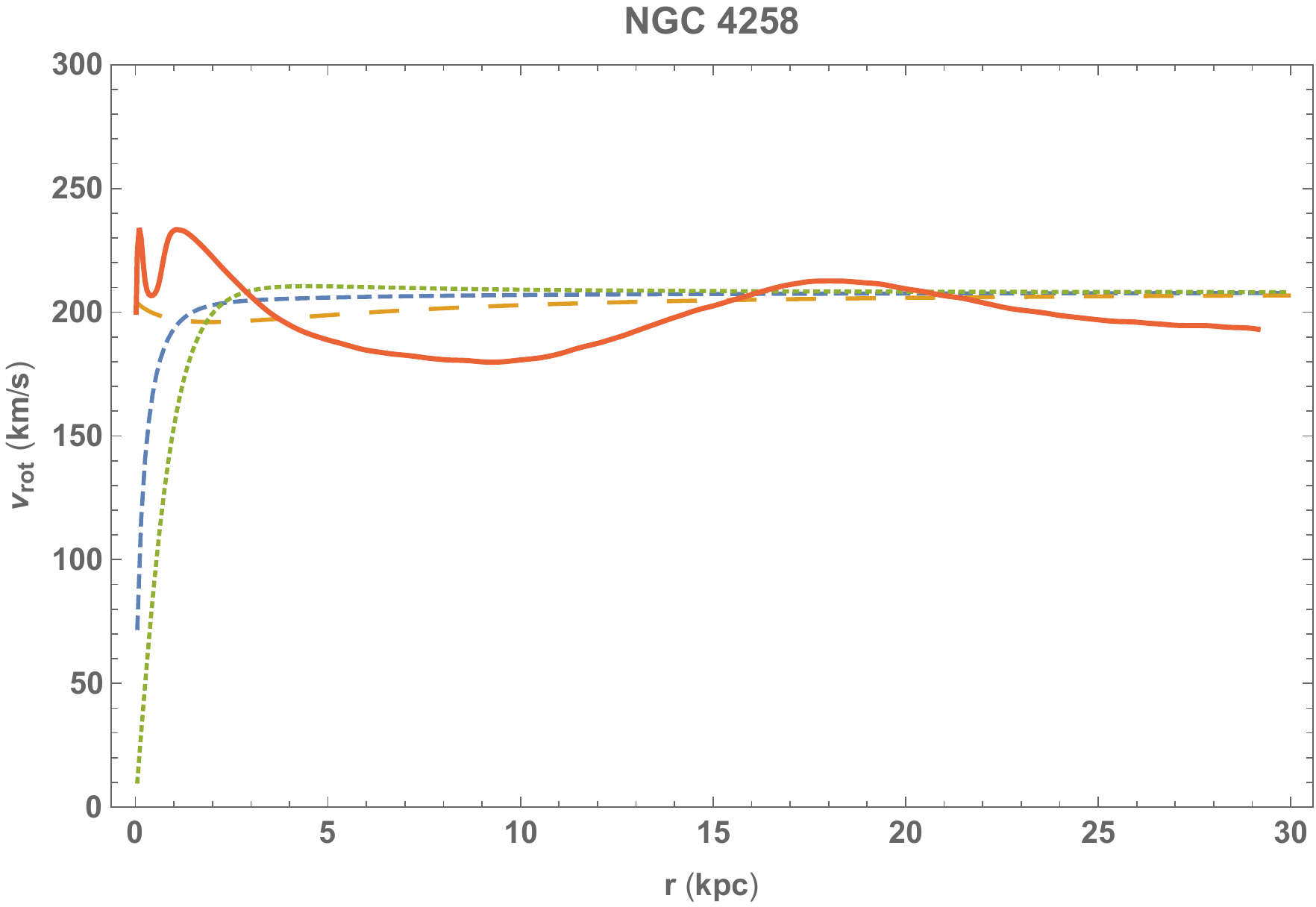}
\nonumber
\end{align}
\caption{Circular velocity as a function of distance for four spiral galaxies : Milky Way, NGC 224, NGC 3079 and NGC 4258. Dataset 1 (with no error bars) for all the four galaxies is from \cite{sofue1999central}  whereas dataset 2 for the Milky Way is taken from \cite{sofue2009unified}. Dataset 3 for NGC 224 is obtained from \cite{braun2009wide}}
\label{fig:rot}
\end{figure}

 The mass densities for DM, and the fraction of the DM mass inside a given radius are shown in Fig. \ref{fig:density-mf}. By construction, the DM mass density exhibits the $ 1/r^2 $ behavior at large $r$ required for the observed flat rotation curves. It is also relatively flat inside the bulge except for the immediate vicinity of the origin where it spikes due to the accumulation of DM around the central black-hole ($\mu$ diverges as $ r \to 0 $, which allows for a higher density of  DM particles to be accommodated in a smaller volume, leading to the observed increase in $ \rho $); though not obvious from the figure, this spike is significant only for $ r \lesssim 1 $ pc. Outside of the region immediately surrounding the black-hole  the exclusion principle obeyed by our DM candidate does  lead to a core-like behavior.   The plot of the DM mass fraction shows that, except for a few kiloparsecs from the galactic center,  galaxies are DM dominated.

In fig \ref{fig:rot} we plot the circular velocity as a function of distance from the galactic center for four spiral galaxies, the Milky Way, NGC 224 (M31 or Andromeda), NGC 3079 and NGC 4258, using the three different baryonic profiles. We also compare the model predictions with data obtained using  CO, HI and H-alpha observations (elliptical galaxies are not included in the sample due to the lack of rotational curve data). The outer region of the rotation curves are in good agreement with the data, as expected from our boundary conditions. The inner dynamics is best reproduced for NGC 3079 followed by the Milky Way, but no so effectively for NGC 224 and NGC 4258. This again, can be attributed to the fact that our model does not include the disc structure, which has a significant contribution to the dynamics of circular velocities, and also assumes complete spherical symmetry for these galaxies. It is then remarkable that the overall qualitative features of the rotation curves for our model are a good fit to the available data.

The statistical errors in the above values for $m$ can be estimated using the scaling relation in \cref{eq:scal.rel}. Using the fact that $ u_0 $ is small for the examples being considered, and taking $ \nu(0)\sim -0.4,\, c(0) \sim 0.9$ (cf. table \ref{fit.params}), we find (at 3 standard deviations)
\beq
 \frac{\delta m}m  \sim 3 \times \left[\half \frac{\delta\rb}\rb -  \frac{\delta \mb}{\mb} +2 \frac{\delta \vrotb}\vrotb \right]
\eeq
assuming that $ \mb \propto a \sigma^2 $  \cite{kormendy2013coevolution}, using \cref{eq:sig-v-spiral,eq:sig-v-ell}, and taking $  \delta \mb/\mb \sim \delta \vrotb/\vrotb \sim 0.1 $  we  find $ \delta m/m \sim 0.4$. This, however, does not include the systematic errors associated with our applying the spherically symmetric model to spiral galaxies, or systematic errors with the data itself; as noted earlier, we expect these errors to be considerably larger.

\subsection{Galaxies without SMBH}
\label{sec:dwarf}

Strong observational evidence suggests that almost all massive galaxies contain a supermassive black hole at their galactic center; most galaxies with no SMBH are small, dwarf galaxies. The best studied members of the latter category are the Milky Way dwarf spheroidal galaxies (dSphs) and because of this, they are the best suited candidates to test our model in the special case where $\mbh=0 $. However, it is widely accepted that these dSphs are mostly dominated by dark matter, with mass-to-light ratios of $M/L_{V} \sim 10^{1-2}$ \cite{mateo1998dwarf}. Detailed studies of light fermionic DM in nearby dwarf spheroidal galaxies have already appeared in the literature \cite{domcke2015dwarf, randall2017cores,destri2013fermionic}, though the implementation of the Thomas-Fermi paradigm is different form the one being discussed here (cf. the discussion in Sect. \ref{sec:introduction} and at the end of Sec. \ref{sec:equil.eq}). The DM profile in our model is determined based on the baryon distribution and hence we do not consider these galaxies due to their negligible baryonic content.

There is also the generally accepted picture that a majority of the dwarf galaxies have slowly rising rotation curves \cite{read2016understanding,swaters2009rotation}, so our assumption of flattened out circular velocities for the boundary conditions no longer holds~\footnote{It is possible to adapt tour approach to these situations, but we will not pursue this here.}. Therefore we will here restrict ourselves to somewhat larger dwarf galaxies without central black holes, but with flat asymptotic rotation curves and also with an estimate of the baryonic mass. We choose a total of eight such dwarf galaxies (from the SPARC database \cite{lelli2016small}) based on their small bulge mass ($\mb \lesssim 10^9 M_\odot$) and small asymptotic rotational velocity ($\vrotb \lesssim 100$ km/s)~\footnote{There were a few other galaxies in the data set that satisfied these  two constraints, but for which we found no real solutions for the DM mass.}. Since we do not find a strong dependence with the baryonic profile function $F$, in this section we restrict ourselves to the case of the Plummer profile. 

\begin{table}[h]
\caption{Dwarf galaxies}
\label{table:dwarf_galaxies}
\begin{tabular}{|c|c|c|} 
 \hline
 \text{Galaxy} &  m (\ev)\\
 \hline
 \text{DDO 154} & 92.8\\
 \text{DDO 168} &  117.6\\
 \text{NGC 2915} & 89.2\\
 \text{NGC 3741} & 114.6\\
 \text{UGC 7603} &  159.5\\
 \text{UGC 5721} & 85.5\\
 \text{UGC 7690} & 65.5\\
 \text{UGC 8550} & 189.2\\

  \hline
\end{tabular}

\end{table}

The values of $m$ for the eight dwarf galaxies are listed in table \ref{table:dwarf_galaxies}; the masses turn out to be on the higher end of the spectrum as compared to the galaxies with SMBHs in the previous section. This can be understood using the scaling relations  \cref{eq:scal.rel,eq:c.nu}, which in this ($u_0 =0 $) case reduces to
\beq
\label{eq:dwarf}
0.412 \ln\left( \frac{\mb}{10^9 M_\odot} \right)+ 0.352 \ln\left( \frac m{30 \ev}\right) = 0.236  \ln\left( \frac \rb{2.5\text{kpc}}\right)  +  0.736 \ln\left( \frac\vrotb{200 \text{km/s}} \right) + 1.493\,,
 \eeq
where we used the fit parameters for the Plummer model listed in table  \ref{fit.params}. For the eight galaxies considered here, if we take the average value of $\vrotb \sim$ 70 km/s and a $\sim$ 0.5 kpc, we get log ($\mb/M_{\odot}$) as 9.17 for the DM mass of 50 eV which is not far off from the data available for $\mb$ (cf. \cite{lelli2016small} ). Also, the farthest outlier in our data, UGC 8550 requires log ($\mb/M_{\odot}$) to be 9.16 as compared to the given value of 8.72. The difference is far less compared to the case of galaxies with SMBH as dwarf galaxy NGC 221 with similar DM mass for the same Plummer profile requires much larger shift in baryonic mass (log ($\mb/M_{\odot}$) of 9.61 as compared to 8.53 provided in the data). This might hint that the large systematic errors in the measurement of $\mb$ are more impactful in the case of galaxies without SMBH causing considerable shift in the DM mass. We again denote by $M'_B$ the total baryon mass when $m$ has the specific value of $50$ eV, then we find that $ M^{\prime}_{B}/\mb$ in the range $1.5-3$  for all the eight dwarfs we studied. As for the case of large galaxies, it is currently impossible to exclude this possibility because of the large systematic errors in $ \mb$.

It should be noted that some of these dwarf galaxies provide two real solutions for the DM mass. In such cases, only the smaller of the two values are included in table \ref{table:dwarf_galaxies} because the  larger mass  solution (in the $\mathcal{O}(500\,\ev)$ range) does not lead to a core-like profile or match with other observations (e.g. rotation curves). 
\begin{figure}[ht]
$$
\begin{array}{lr}
\includegraphics[width=2.3in]{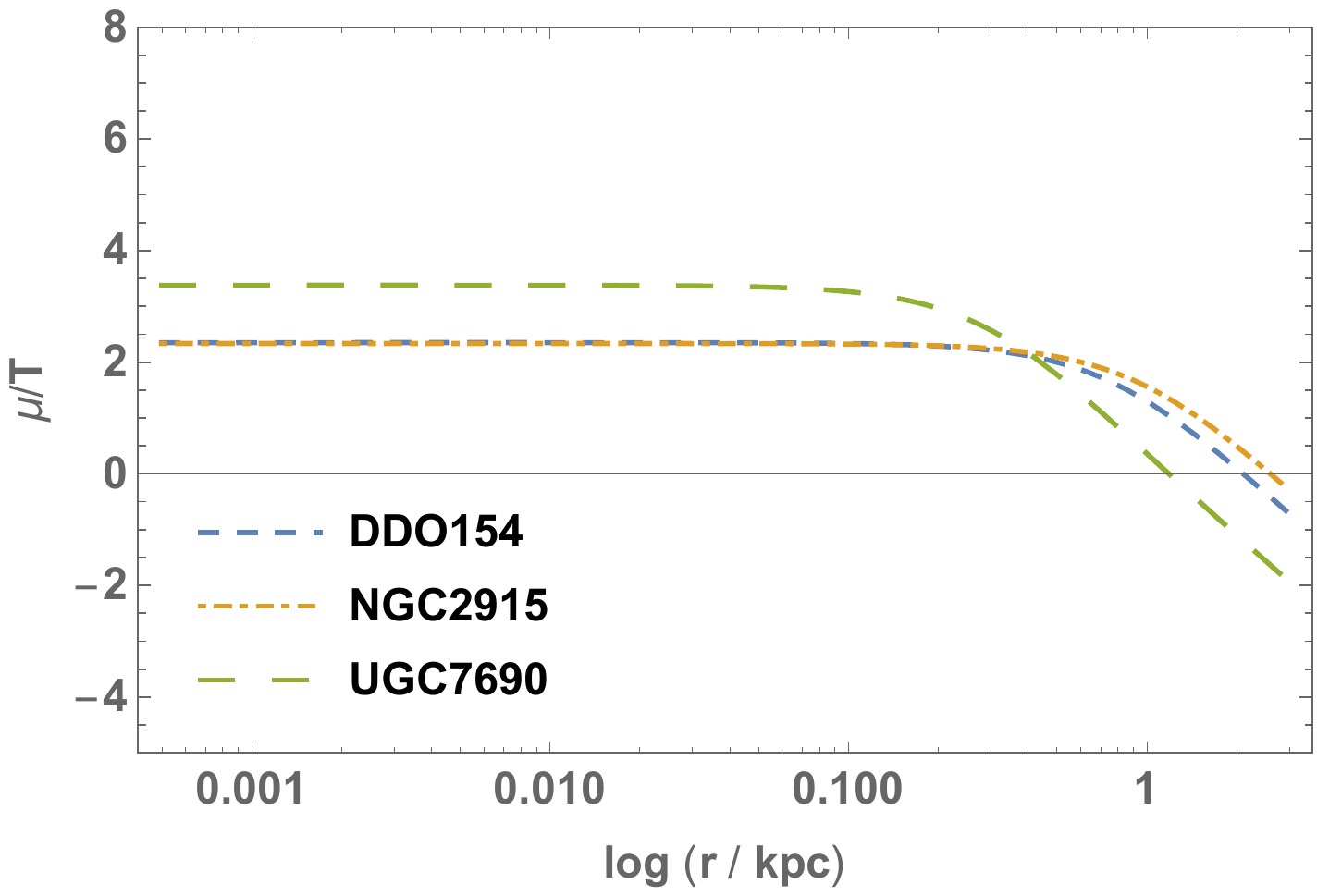} & \qquad
\includegraphics[width=2.3in]{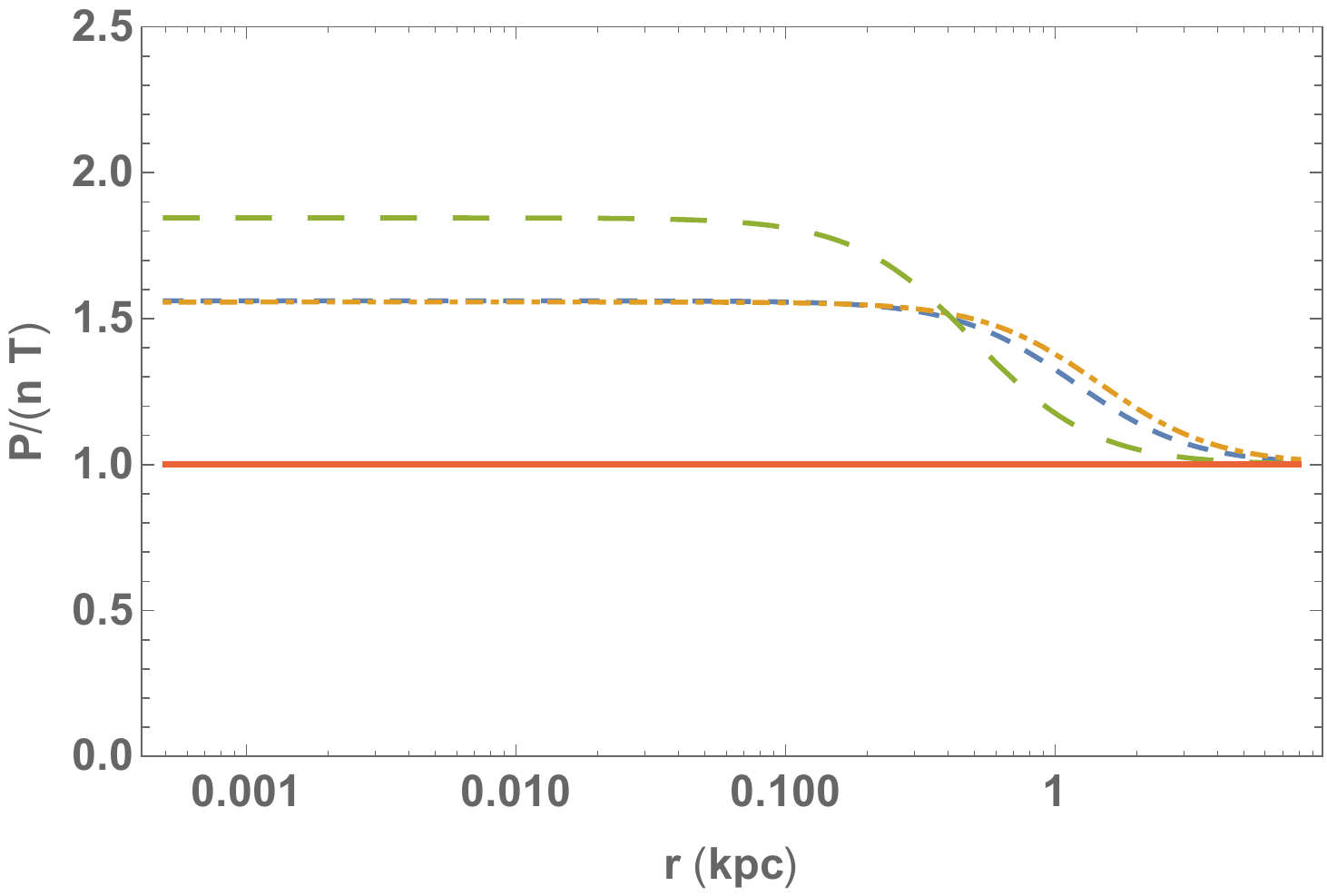} \cr
\includegraphics[width=2.3in]{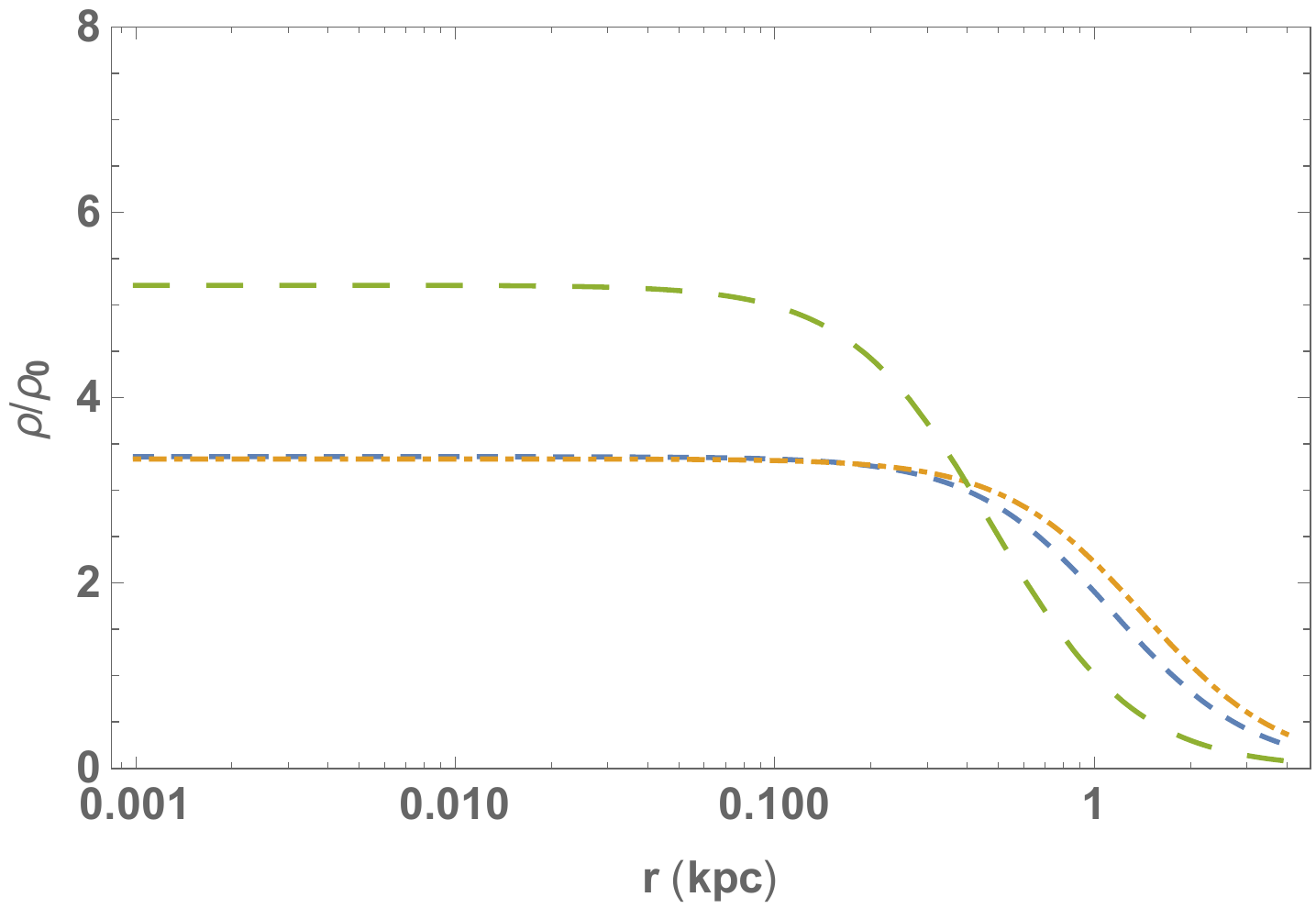} & \qquad
\includegraphics[width=2.3in]{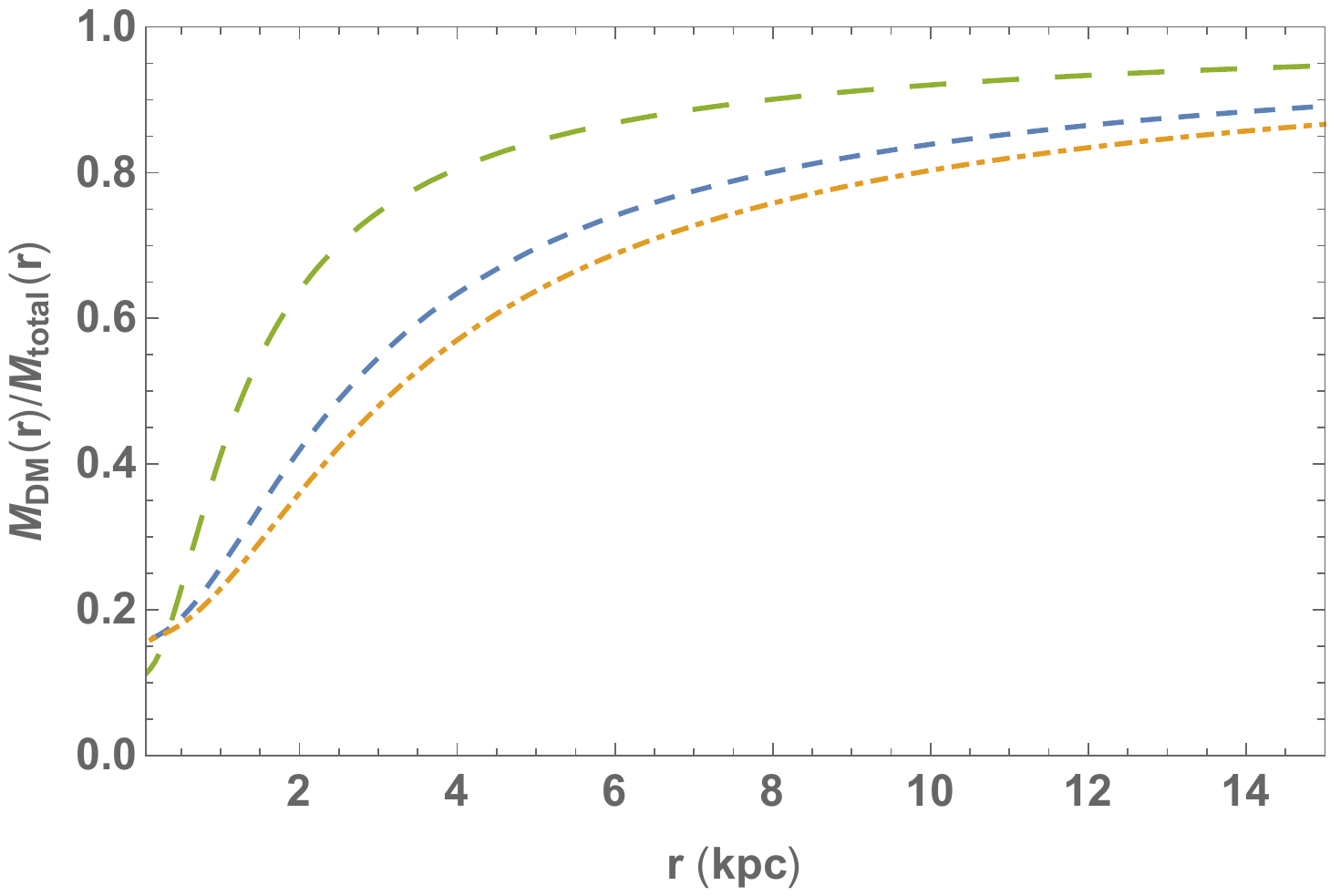} \cr
\includegraphics[width=2.3in]{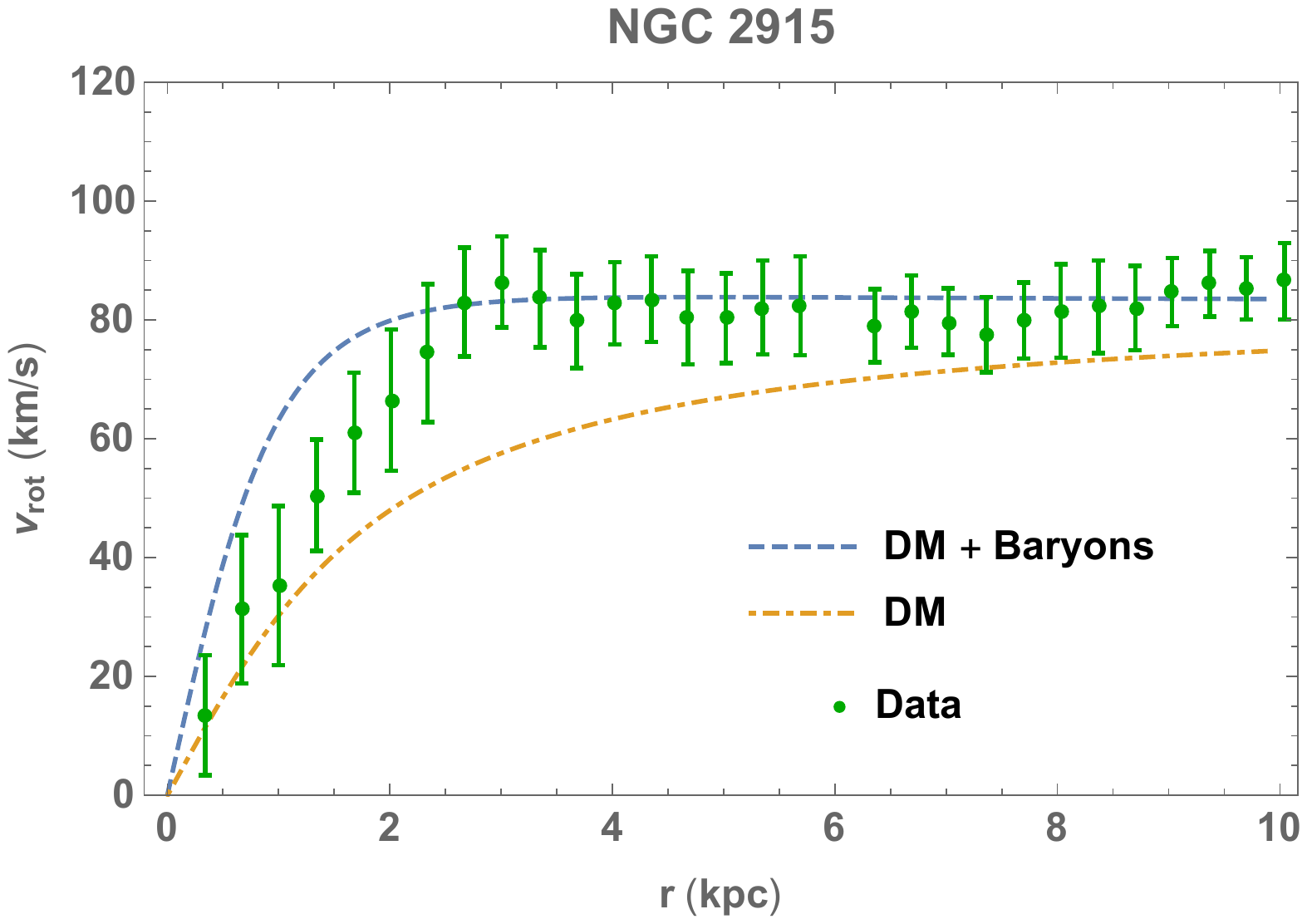} & \qquad
\includegraphics[width=2.3in]{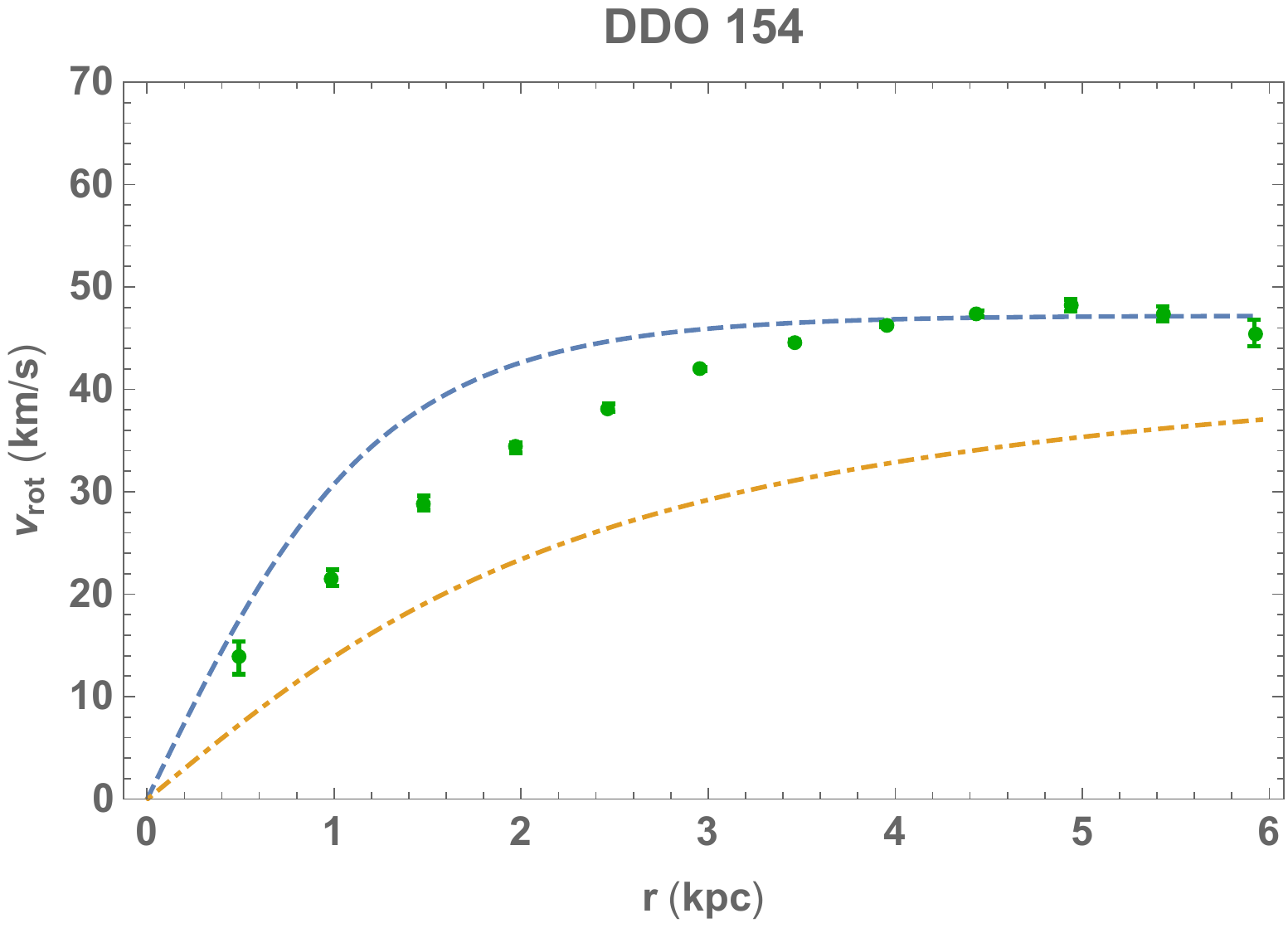}
\end{array} $$
	\caption{\footnotesize Properties of the solution to the TFDM equations for dwarf galaxies. Top row: chemical potential (left) and $P/(nT) $ (right) for the DM as a function of $r$ for 3 dwarf galaxies; middle row: DM density (left) and mass fraction (right) for the same galaxies; bottom row: rotation curve for NGC 2915 (left) and DDO 154 (right) with rotation data taken from \cite{lelli2016sparc}; also, $ \rho_0 = 2 m/\lambda^3$.}
\label{fig:dwarf_props}
\end{figure}

In Fig \ref{fig:dwarf_props}  we illustrate the properties of the solutions by plotting various properties of model predictions for three dwarf galaxies, whose behavior away from the center is qualitatively similar to that of large galaxies with SMBHs. We note that the predicted dark matter profiles show a central constant density core with core radii $r \sim 100$-$400$ pc (which is also the case for the other galaxies in our set). Of special interest are the rotation curves (bottom line in the figure): for DDO 154 and NGC 2915, the predicted behavior of $\vrot(r)$ qualitatively matches quite well with the observations, but the rise in the curve is somewhat steeper compared to the data. It is unclear whether these discrepancies are due to a shortcoming in the model itself or in the simplifying assumptions we adopted, or due to the specific baryonic profile (Plummer's) we use~\footnote{The match with observations does not improve if we use the Hernquist or Jaffe profiles.}.

\section{Conclusions}
\label{sec:conclusions}
In this paper we investigated the extent to which a DM model consisting of single, light fermion, is consistent with the observed bulk properties of galaxies (effective radius, baryon mass and profile, etc.). To simplify the calculations we neglected possible fermion (non-gravitational) interactions, and assumed that the galaxies are well described by a spherically-symmetric configuration. We also assumed a fixed baryon distribution that affects the mechanical equilibrium of the system, but we neglected any thermal  or dynamical effects of the baryons. The baryon profile, which is directly observable, together with the boundary conditions leading to flat rotation curves, completely determine the DM distribution in the system. This is in contrast with other publications which assume a DM profile {\it ab initio}.

For the set of galaxies we considered (that includes spiral, ellipticals and several dwarf galaxies) the model is consistent with the observational data, in the sense that the values of $m$ we obtain lie in a relatively narrow range. Admittedly, for the model to be convincing, the {\em same} value of $m$ should be obtained for all galaxies; but to test this  would require a careful modeling of each galaxy, and solving the stability equation \cref{eq:eom} without the assumption of spherical symmetry -- which lies beyond the scope of this paper. A stringent test of the model would also require more accurate data with reduced systematic errors; it is unclear whether any of these effects leads to the $ m - \mb$ correlation observed in Fig. \ref{fig:dMB-MB}. Given these uncertainties we limit ourselves to stating that the model is promising, but additional calculations and observations are necessary to fully determine its viability.

For galaxies with a SMBH we find that the preferred DM mass is $ \sim 40 $ eV, and that the DM distribution has a central core region where the fermions are strongly degenerate, with the degeneracy increasing as the central black-hole is approached. For galaxies without SMBHs the DM mass values we find are generally larger ($ \gtrsim 70 $ eV). Possible reasons for this discrepancy, as well as for the spread in the preferred values of $m$ within each galaxy class are discussed in sections \ref{sec:large} and \ref{sec:dwarf}. It is interesting to note that the lower bounds for $m$ obtained in \cite{randall2017cores, domcke2015dwarf,di2018phase} for the Milky Way dwarf spheroidal galaxies are in the range $20-100$ eV, which is consistent with our results for galaxies without a SMBH. 

Interestingly, this model makes clear testable predictions that may be worth exploring in more detail. For instance, at fixed $m$ and asymptotic outer velocity, the profile is fully determined by the equilibrium reached between dark matter and baryons. This means that any detected difference in the {\it shapes} of the rotation curves measured in galaxies at fixed terminal rotation velocity \cite{Oman2015}, in particular for dark matter-dominated objects like dwarfs, should be accompanied by a significant difference in the baryonic mass distribution. Such correlation has already been shown to help alleviate the problem of rotation velocity diversity in the case of self-interactive dark matter \cite{Creasey2017}. Exploring the correlation between observed baryonic properties (mass, gas fractions, size) and the shape of the velocity profiles in single fermion dark matter case  would also help assess the viability of this model.

Small deviations from spherical symmetry can be implemented using perturbation theory, which would be applicable to elliptical galaxies or for studying the effects of rotation. In contrast, a more accurate comparison of the model to spiral galaxies will require solving \cref{eq:eom} assuming cylindrical symmetry, and including in  $ \rho_B$ bulge and spiral components. Also of interest would be a study of the dynamic stability of the system, that can be approached using standard techniques \cite{siegel1971}; in this case \cref{eq:eom} is replaced by the Euler equation and complemented by the DM and baryon current conservation constraints.

Finally, we wish to comment on the possible effects of exchange interactions. Inside an atom these effects are significant \cite{Bethe-Jackiw}, but in the present situation they can be neglected since we assume the fermions experience only gravitational interactions. This, however, will change dramatically should fermion self-interactions are included, and can lead to a further reduction of the DM pile-up at the core.

\begin{acknowledgments}
The authors would like to thank Hai-bo Yu for interesting and useful comments. LVS acknowledges support from NASA through the HST Program AR-14582 and from the Hellman Foundation.
\end{acknowledgments}

\appendix
\section{Comments on the data used.}
\label{appendix:data}
In this appendix we give some details on the data we used to obtain the results presented in the main text.

For galaxies with SMBHs, we consider in total a sample of 60 galaxies, 29  elliptical and 31  spiral galaxies. For each of these galaxies, we needed the mass of the black hole  $\mbh$, bulge mass $\mb$, scale radius $\rb$ and the asymptotic velocity $\vrotb$.
We got most of the entries in our dataset from \cite{hu2009black} (we used $\mb$ calculated by K band M/L derived from B-V color, and excluded galaxies where this value of $\mb$ was unavailable). In addition, we obtained $\mbh,\,\mb$ and $ \vrotb$ from \cite{kormendy2013coevolution} for 3 elliptical (NGC 1332,  NGC 1407 and NGC 7052 ) and 2 spiral galaxies (NGC 1277 and NGC 3945); for these 5 galaxies we obtained $ \rb$ from 3 sources: \cite{lelli2017one} for NGC 1332, NGC 1407 and NGC 3945; \cite{hu2009black} for NGC 7052 and  \cite{graham2016normal} for NGC 1277. Other galaxies from \cite{kormendy2013coevolution} were not included due to the lack of data on the effective/half-light radius.

The asymptotic circular $\vrotb$ for some of the spiral galaxies (Circinus, Milky Way, NGC 224, NGC 1023, NGC 1068, NGC 2787, NGC 3031, NGC 3115, NGC 3227, NGC 3384, NGC 3585, NGC 4026, NGC 4258,  NGC 4596, NGC 7457 and  IC2560)  are listed in \cite{kormendy2013coevolution}. For all other spiral galaxies we use the empirical relation  \cite{ho2007bulge},
\beq
 \log \vrotb = (0.8 \pm 0.029) \log\sigma  + (0.62 \pm 0.062)\,,
 \label{eq:sig-v-spiral}
 \eeq
where $\sigma$ is the bulge velocity dispersion. For elliptical galaxies, we assume a very similar relation from the same reference:
\beq
\log\vrotb = (0.82 \pm 0.027) \log\sigma  + (0.57 \pm 0.058),
\label{eq:sig-v-ell}
\eeq  
that was obtained using a larger sample of galaxies  including ellipticals. The data for rotation curves of spiral galaxies is taken from \cite{sofue1999central} . 

For galaxies with no central black hole, we obtained $ \mb,\,\rb$ and $ \vrotb$ from the SPARC database \cite{lelli2016sparc, lelli2016small}. We note that this dataset has no information on the presence or absence of SMBHs, so we include only eight of the smallest dwarf galaxies. 

\bibliographystyle{apsrev4-1}
\bibliography{ULTFDM}

\end{document}